\documentclass[conference]{IEEEtran}
\usepackage[utf8]{inputenc}
\usepackage{algpseudocode}
\usepackage[ruled,vlined,linesnumbered]{algorithm2e}
\usepackage{varwidth}
\usepackage{epigraph}
\usepackage{graphicx}
\usepackage{listings}
\usepackage{textcomp}
\usepackage{xcolor}
\usepackage{url}
\usepackage{subcaption}
\usepackage{enumerate}
\usepackage{xspace}
\usepackage{xcolor}
\usepackage{booktabs}
\usepackage{multirow}
\usepackage{mathtools}
\usepackage{forest}
\usepackage{tikz}
\usepackage{tikz-qtree}
\usetikzlibrary{arrows.meta, decorations.pathreplacing, shadows, positioning}
\usepackage[binary-units,mode=text]{siunitx}
\usepackage{paralist}
\usepackage{wasysym}
\usepackage{boldline}
\usepackage[all]{nowidow}
\usepackage{flushend}

\usepackage[capitalize]{cleveref} 

\newcommand{\name}{F-PKI\xspace}

\DeclarePairedDelimiter\ceil{\lceil}{\rceil}

\renewcommand{\paragraph}[1]{\noindent\textbf{#1}\quad}

\newcommand{\ys}{{$\vcenter{\hbox{\scriptsize\CIRCLE}}$}\xspace} 
\newcommand{\no}{{$\vcenter{\hbox{\scriptsize\Circle}}$}\xspace} 
\newcommand{\pt}{{$\vcenter{\hbox{\scriptsize\LEFTcircle}}$}\xspace} 

\begin{document}

\title{\name: Enabling Innovation and Trust Flexibility in the HTTPS Public-Key Infrastructure}

\author{\IEEEauthorblockN{Laurent Chuat}
\IEEEauthorblockA{ETH Zurich}
\and
\IEEEauthorblockN{Cyrill Krähenbühl}
\IEEEauthorblockA{ETH Zurich}
\and
\IEEEauthorblockN{Prateek Mittal}
\IEEEauthorblockA{Princeton University}
\and
\IEEEauthorblockN{Adrian Perrig}
\IEEEauthorblockA{ETH Zurich}}

\maketitle

\begin{abstract}
We present \name{}, an enhancement to the HTTPS public-key infrastructure (or web PKI) that gives
trust flexibility to both clients and domain owners, and enables certification authorities (CAs) to
enforce stronger security measures. In today's web PKI, all CAs are equally trusted, and security is
defined by the weakest link. We address this problem by introducing trust flexibility in two
dimensions: with \name{}, each domain owner can define a domain policy (specifying, for example,
which CAs are authorized to issue certificates for their domain name) and each client can set or
choose a validation policy based on trust levels. \name{} thus supports a property that is sorely
needed in today's Internet: trust heterogeneity. Different parties can express different trust
preferences while still being able to verify all certificates. In contrast, today's web PKI only
allows clients to fully distrust suspicious/misbehaving CAs, which is likely to cause collateral
damage in the form of legitimate certificates being rejected. Our contribution is to present a
system that is backward compatible, provides sensible security properties to both clients and domain
owners, ensures the verifiability of all certificates, and prevents downgrade attacks. Furthermore,
\name{} provides a ground for innovation, as it gives CAs an incentive to deploy new security
measures to attract more customers, without having these measures undercut by vulnerable CAs.
\end{abstract}

\section{Introduction}
\label{sec:introduction}

In March 2011, news broke that Comodo---a security firm operating a certification authority---had
been hacked. The intrusion resulted in the unwarranted issuance of 9 certificates for several
high-profile domain names~\cite{comodo}. A few months later, DigiNotar suffered a similar
attack~\cite{hoogstraaten2012black}. These events led Google to create the Certificate Transparency
(CT) framework~\cite{rfc6962}. About 8 years later, CT is in the final stages of its
deployment~\cite{Stark2019CT}. Transparency greatly facilitates the \emph{detection} of illegitimate
certificates, but there remain the questions of how to \emph{react} after misbehavior is observed
and how to \emph{prevent} misbehavior altogether. Unfortunately, simply revoking the certificates of
vulnerable CAs would have serious consequences: all the certificates issued by these CAs would
become invalid, rendering countless websites unavailable. An ideal public-key infrastructure would
prevent a vulnerable or misbehaving CA from jeopardizing the security of the entire system in the
first place, and it would give users and browser vendors an option to demote CAs without completely
distrusting them. Unfortunately, the definition of trust in traditional PKIs is too rigid to enable
this ideal vision. 

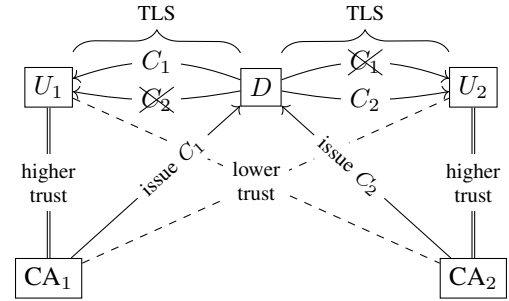
\begin{figure}
	\centering
	\begin{tikzpicture}[squarednode/.style={rectangle, draw=black, minimum size=5mm}, roundedcornernode/.style={rectangle, draw=black, rounded corners}]
          \node[squarednode] (u1) {$U_1$};

          \node[squarednode] (u2) [right = 50mm of u1] {$U_2$};

          \node[squarednode] (d) at ($(u1)!0.5!(u2)$) {$D$};
          \draw [->] (d) to [bend left=10] node[fill=white] (u1c2) [midway] {$C_2$} (u1);
          \draw [->] (d) to [bend right=20] node[fill=white] (u1c1) [midway] {$C_1$} (u1);
          \draw ($(u1c2.north west)+(1mm,-1mm)$) to ($(u1c2.south east)+(-1mm,1mm)$);
          \draw ($(u1c2.north east)+(-1mm,-1mm)$) to ($(u1c2.south west)+(1mm,1mm)$);
          \draw [decorate,decoration={brace,amplitude=3mm,raise=2mm}] (u1.north east) -- (d.north west) node [black,midway,above,anchor=south,yshift=5mm] {\footnotesize TLS};
          \draw [->] (d) to [bend left=20] node[fill=white] (u2c1) [midway] {$C_1$} (u2);
          \draw [->] (d) to [bend right=10] node[fill=white] (u2c2) [midway] {$C_2$} (u2);
          \draw ($(u2c1.north west)+(1mm,-1mm)$) to ($(u2c1.south east)+(-1mm,1mm)$);
          \draw ($(u2c1.north east)+(-1mm,-1mm)$) to ($(u2c1.south west)+(1mm,1mm)$);
          \draw [decorate,decoration={brace,amplitude=3mm,raise=2mm}] (d.north east) -- (u2.north west) node [black,midway,above,anchor=south,yshift=5mm] {\footnotesize TLS};

          \node[squarednode] (ca1) [below = 20mm of u1] {$\text{CA}_1$};
          \draw[double] (u1) to node[fill=white,align=center,font=\footnotesize] [midway] {higher\\trust} (ca1);
          \draw[dashed] (ca1) to (u2);

          \node[squarednode] (ca2) at (u2 |- ca1) {$\text{CA}_2$};
          \draw[dashed] (u1) to node[fill=white,align=center,font=\footnotesize] [pos=0.5] {lower\\trust} (ca2);
          \draw[double] (u2) to node[fill=white,align=center,font=\footnotesize] [midway] {higher\\trust} (ca2);

          \draw[->] (ca1) to node[fill=white,align=center,font=\footnotesize,sloped] [pos=0.6] {issue $C_1$} (d);
          \draw[->] (ca2) to node[fill=white,align=center,font=\footnotesize,sloped] [pos=0.6] {issue $C_2$} (d);
  	\end{tikzpicture}
	\caption{A dashed line indicates a lower trust and a double line indicates a higher trust.
          An arrow indicates certificate issuance or a certificate sent during a TLS handshake.}
	\label{fig:trust-incompatibility}
\end{figure}

Our central observation is that trust is highly heterogeneous across the world. A PKI that supports
trust flexibility would provide a foundation for domain owners and users/browsers to express trust
preferences, penalize misbehaving or vulnerable CAs, and reward CAs with strong security measures.
The core challenges in enabling heterogeneous levels of trust in CAs are to achieve a meaningful
overall system behavior with concrete security properties, ensure global verifiability of all
certificates, and prevent downgrade attacks to lower security levels. An example to illustrate this
is shown in Figure~\ref{fig:trust-incompatibility}. In this example, user $U_1$ trusts $\text{CA}_1$
more than $\text{CA}_2$ for issuing certificates for domain $D$ because $\text{CA}_1$ supports
multi-perspective domain validation~\cite{MultiPerspectives}, while user $U_2$ trusts $\text{CA}_2$
more than $\text{CA}_1$ because $\text{CA}_2$ is an American CA and $D$'s TLD is .us. In
this example, $U_1$ should be able to express higher trust in $\text{CA}_1$ than in $\text{CA}_2$,
while retaining the ability to use certificates issued by $\text{CA}_2$. A challenge immediately
arises: how can such a policy result in a clear security property for the user? We resolve
this challenge as follows: the user (or browser vendor) may define a set of CAs as highly trusted
for a set of domains, then the browser obtains assurance that a received certificate does not
conflict with any certificate issued by a highly trusted CA. In this way, from each user's
perspective, a clear and consistent security guarantee is provided.

\name{} (which stands for flexible public-key infrastructure) introduces trust flexibility in
two dimensions: each domain can set a domain policy and
each verifier can set (or choose) a validation policy using trust levels for CAs. Domain owners can
specify policies in their certificates to restrict the set of valid certificates for their domain.
Clients are then presented with a comprehensive set of certificates for each domain they visit and
can make informed decisions based on their validation policy. \name{} allows these clients to
express a preference for certain CAs. Our new notion of trust is ternary and name-dependent: every
authority may be either untrusted, trusted, or highly trusted, for each domain name. In our example,
user $U_1$ would treat $\text{CA}_1$ as highly trusted and $\text{CA}_2$ as trusted for domain $D$.
$U_1$ would then reject $C_2$ if it conflicts with $C_1$. We envision that browser vendors would
initially dictate trust levels, but users could modify these default trust levels as they please.
Once trust levels are defined, a user should only accept a certificate if it comes with evidence
that no highly trusted CA has issued a conflicting certificate.

Domain owners are given the option to define what constitutes a ``conflicting'' certificate for
their domain, as \name{} gives them the ability to specify policies through certificate extensions.
Users then receive and consider all policies signed by highly trusted CAs. The \emph{issuers}
policy, for example, lists CAs authorized to issue certificates for a domain name. In our example,
if the owner of $D$ sets such a policy in $C_1$, $U_1$ will reject $C_2$ if $\text{CA}_2$ is not
listed as an issuer in $C_1$. Giving domain owners the ability to define policies through
certificate extensions would be futile, however, if an attacker who is able to obtain a bogus
certificate could simply hide those policies. In our example, if $U_1$ does not know about the
policies defined in $C_1$, then $U_1$ would accept $C_2$. Therefore, we introduce a verifiable log
server that can provide users with a view of all certificates and revocation messages relevant to
any given domain name. This new log server, which we call map server, is meant to complement
existing Certificate Transparency servers.

\paragraph{Main Contributions.}
We introduce a new trust model for the web PKI, which takes into account the heterogeneity of trust
around the world: each user can assign each CA to a different trust level for each domain name.  We
present \name{}, which allows users to make informed decisions when validating certificates by
considering a global set of certificates rather than a single certificate chain.  \name{} prevents
downgrade attacks to a less trusted CA, while retaining verifiability of all certificates.  Any
domain owner can opt-in to obtain additional security guarantees simply by requesting a new
certificate with an appropriate X.509 extension, which preserves backward compatibility to the
existing web PKI\@.  We demonstrate that \name{} can be realized in practice and deployed
incrementally by upgrading CT log servers or by introducing a new verifiable log server (called map
server).  \name{} does not require server updates and does not require active CA participation
(beyond support for certificate extensions).  Additionally, \name{} incentivizes CAs to innovate and
offer new security measures, as it prevents other (vulnerable) CAs from undercutting these measures.
Finally, we present a proof-of-concept implementation and evaluate it to show that \name{} is
capable of withstanding realistic workloads and prevent attacks with low overhead.

\section{Background: Verifiable Logging}

\paragraph{Certificate Transparency (CT).}
The objective of CT is to log certificates and make them publicly available, so that misbehavior can
be detected. Merkle hash trees facilitate the audit of log servers. In a Merkle hash tree (MHT, also
referred to as Merkle tree or hash tree), leaves contain data, while intermediate nodes and the root
of the tree contain the hash of concatenated child nodes. Typically, Merkle trees are binary so two
children are concatenated and hashed to determine the value of their parent node. By appending new
entries to a Merkle tree in chronological order, CT logs can efficiently produce presence and
consistency proofs. Proving the presence of a leaf in a Merkle tree whose root is known is
efficient: a number of nodes (logarithmic in the number of leaves in the tree, assuming a
balanced tree) are provided to the verifier so that they can reconstruct the path to the root. The
role of a consistency proof, on the other hand, is to corroborate the supposed append-only property
of CT logs. Again, a number of nodes logarithmic in the number of leaves is sufficient to prove
consistency between two versions of a log, assuming entries are added chronologically.

\paragraph{Absence proofs.}
The absence of an entry cannot be efficiently proven using a chronological hash tree. Absence
proofs, although not required by CT, are necessary in other contexts. Laurie and
Kasper~\cite{RevTrans} proposed to support absence proofs for \emph{revocation transparency} using a
sparse Merkle tree of intractable size. In such a tree, each output of a hash function has a
distinct leaf; using SHA256, the tree has $2^{256}$ leaves. It is possible to use a sparse Merkle
tree in practice (despite its intractable size), because most leaves are empty and thus most nodes
have the same predictable values. This fact can be exploited to make sparse Merkle trees efficient
with caching strategies~\cite{dahlberg2016efficient}. Trillian~\cite{Trillian} is an open-source
verifiable data store, implemented in Go and developed at Google. The white
paper~\cite{VerifDataStruct} describing the underlying data structures of Trillian expands upon the
idea of sparse Merkle trees. Trillian offers three modes of operation: (a) verifiable log
(equivalent to a CT log), (b) verifiable map (equivalent to a sparse Merkle tree), and (c)
verifiable log-backed map (which uses a combination of both tree types).

\section{Lessons Learned}\label{sec:lessons-learned}

Certificate Transparency has been a tremendous success, but it does not enable proactive security
measures; it only allows detecting misbehavior after the fact. If a CA is compromised or
malicious, detecting illegitimate certificates within an unspecified timeframe is not sufficient.
The ability to issue certificates for any domain combined with a man-in-the-middle attack can be
devastating, as both the confidentiality and integrity of all web communications are threatened.
Mechanisms aimed at preventing such attacks have been proposed in the past, but deploying them
has been a challenge (sometimes with limited benefits).

HTTP Public Key Pinning~(HPKP), which is now deprecated by major browsers~\cite{FirefoxHPKP}, was
designed to fulfill an objective similar to ours, i.e., to prevent an attacker from using an
illegitimate certificate when another public key is already bound to the domain name in question. In
a nutshell, HPKP works as follows: A web server sends a pinning policy to a client through an HTTP
header field. The policy may specify a public key that the client should expect to find in the
server's certificate. The client then enforces the received policy for each connection to the
domain. The ``max-age'' directive specifies the time during which the policy should be
enforced~\cite{rfc7469}. Unfortunately, HPKP can easily be misused, by attackers and domain owners
themselves~\cite{Helme2016}. An attacker can launch a ``ransom~PKP'' attack by pinning a public key,
before asking the owner for a ransom in exchange for a private key whose public counterpart was
pinned during the attack. A domain owner can also inadvertently commit ``HPKP Suicide'' by pinning a
public key without knowing the corresponding private key. Moreover, HPKP is only effective after the
first connection is established and for a limited amount of time. A number of other mechanisms, such
as HSTS~\cite{rfc6797} and OCSP Must-Staple~\cite{rfc7633}, suffer from the same problem. Besides,
HSTS only enforces the use of HTTPS, while OCSP stapling only addresses the revocation problem;
neither provides resilience against CA compromise. Using a different approach, CAA records and
DANE~\cite{rfc7671} were proposed to address the problem of misbehaving CAs, but both rely on
DNSSEC, which requires its own PKI and also suffers from deployment issues. Moreover, CAA is only
intended for CAs: a rogue CA can completely ignore CAA policies, while our system is designed
to prevent misbehavior by letting clients verify policies on their own.

The lessons we can learn from these schemes are the following:
(a) The infrastructure itself should be designed to support domain policies.
(b) It should not be possible for a domain owner to pin a public key to their domain name, unless
    they can prove possession of the corresponding private key.
	For this reason, our policies will be defined through certificate extensions.
(c) Once a policy is advertised to some clients, it should be possible to revoke it.
(d) PKI improvements should not require updating all web servers. CT has shown that CAs are more
    likely to adopt a new security scheme (if they have incentives to do so) than individual domain
    owners.
(e) The security of the web PKI should not rely on trust on first use (TOFU) or on the security of
    a separate infrastructure.
    
\Cref{tab:comparison} shows how \name{} compares with PKI schemes and enhancements that were (at
least partially) deployed and supported by browsers. Our comparison is inspired by previous similar
analyses~\cite{BCKPSS2016,CT_comparison}.  We discuss more related work in \cref{sec:related}.
Below are the criteria we used in our comparison:

\begin{itemize}
	\item \textbf{Prevents attacks:} If the web server is replaced by a malicious server, the scheme
	will prevent anyone from connecting to that server.
	\item \textbf{Detects global attacks:} If the web server is replaced by a malicious server that
	everyone sees, the scheme will help the domain owner detect the attack.
	\item \textbf{Detects targeted attacks:} If the web server is replaced by a malicious server
	that only a small number of people can see, the scheme will still help the domain owner detect
	the attack.
	\item \textbf{Built-in revocation:} the scheme supports some form of certificate revocation.
	\item \textbf{Unmodified server:} web servers do not have to be modified to support the scheme
    and still benefit from additional security guarantees.
    \item \textbf{Instant startup:} a new web server can use a certificate and be trusted by clients
	immediately.
	\item \textbf{Instant recovery:} if the private key is lost, a new certificate can immediately
	be used.
	\item \textbf{No out-of-band communication:} no side channel is required to support the scheme.
	This is only partially the case for F-PKI because clients must contact a map server; however,
	the connection to the map server can be established via a regular DNS resolver and before the
	TLS session is established. Similarly, DANE and CAA rely on DNSSEC.
	\item \textbf{No log synchronization required:} If the scheme uses (or supports the use of)
	multiple logs, they do not need to be synchronized. Although a map server could be replicated
	with an appropriate consensus protocol~\cite{logres2020}, F-PKI can work with a single map
	server or several unsynchronized map servers.
	\item \textbf{Supports multiple certificates:} several certificates can be used simultaneously
	for the same domain.
\end{itemize}

\begin{table}[t]
    \centering
    \small
    \begin{tabular}{ l | *{11}{c@{\hspace{.5em}}} }
        &
        \rotatebox[origin=lB]{90}{\textbf{Prevents attacks}} &
        \rotatebox[origin=lB]{90}{\textbf{Detects global attacks}} &
        \rotatebox[origin=lB]{90}{\textbf{Detects targeted attacks}} &
        \rotatebox[origin=lB]{90}{\textbf{Built-in revocation}} &
        \rotatebox[origin=lB]{90}{\textbf{Unmodified server}} &
        \rotatebox[origin=lB]{90}{\textbf{Instant startup}} &
        \rotatebox[origin=lB]{90}{\textbf{Instant recovery}} &
        \rotatebox[origin=lB]{90}{\textbf{No out-of-band comm.}} &
        \rotatebox[origin=lB]{90}{\textbf{No log sync. required}} &
        \rotatebox[origin=lB]{90}{\textbf{Supports multiple cert.}} \\
        \hlineB{2} \\[-2ex]
        CT	 		& \no & \ys & \ys & \no & \ys & \ys & \ys & \ys & \ys & \ys \\
		DANE		& \ys & \ys & \ys & \no & \pt & \ys & \ys & \pt & \ys & \ys \\
		CAA			& \no & \no & \no & \no & \pt & \ys & \ys & \pt & \ys & \ys \\
		HPKP		& \pt & \pt & \pt & \no & \no & \pt & \no & \ys & \ys & \ys \\
		F-PKI$^1$	& \ys & \ys & \ys & \ys & \ys & \ys & \ys & \pt & \ys & \ys \\
		F-PKI$^2$	& \ys & \ys & \ys & \ys & \no & \ys & \ys & \ys & \ys & \ys \\
		\hlineB{2}
    \end{tabular} \\[1ex]
    \ys Yes \qquad \pt Partially \qquad \no No \\[1ex]
    \caption{Comparison of PKI schemes/enhancements. All criteria are expressed as benefits
    (i.e., positively). $^1$DNS-based deployment $^2$Stapling-based deployment}
	\label{tab:comparison}
\end{table}

\section{Trust Model}
\label{sec:model}

In this section, we introduce a new, more flexible trust model for the web PKI\@. Let the
\emph{relying party} be any entity (e.g., a client) that uses the public key in a
certificate~\cite{rfc3647}. Our primary goal is to prevent a CA from attacking a domain if a
certificate has already been issued by another CA that the relying party trusts more for the domain
name in question. To achieve this goal, we extend the trust model of today's web PKI in two
ways. First, we introduce a new trust level: each relying party may consider some CAs more trusted
than others, for all names or a subset thereof. In other words, trust in our PKI is now
\emph{ternary} and \emph{name-dependent}.  Second, we introduce new domain policies, which are set
by the respective domain owner and are certified by a CA alongside the certificate.  The relying
party considers all domain policies certified by highly trusted CAs.
As opposed to X.509 name constraints~\cite{rfc4158}, all CAs can still issue certificates for any
domain as long as they don't interfere with the certificates issued by highly trusted CAs.
The three trust classes of our model are the following:

\noindent
\textbf{Untrusted:} As in today's trust model, only a public key that is part of a valid CA
certificate can be used to verify certificate signatures. Other public keys are untrusted.

\noindent
\textbf{Standard trust (non-highly trusted):} This corresponds to the current notion of trust in a
CA. Any CA in this trust class can keep issuing certificates, as in today's web PKI.  However if an
issued certificate violates a policy defined by a highly trusted CA, it is rejected by the relying
party.

\noindent
\textbf{Priority trust (highly trusted):} This is the new trust level we introduce. Some CAs may
be highly trusted for a set of names. Relying parties only consider the policies defined by CAs
that they highly trust. This is defined by the following function:

\smallskip
\centerline{$f(N)$: set of authorities highly trusted for name $N$.}
\smallskip

This model could be extended to support more trust classes. However, the number of trust 
classes should be kept low (for usability reasons), while still allowing relying parties to
distinguish between reputable, neutral, and untrustworthy CAs.

A more formal representation of this trust model is presented in \cref{sec:formal_model}.

\subsection{Adversary Model}
\label{sec:adversary_model}

We assume that the attacker's capabilities are constrained as in the Dolev--Yao
model~\cite{Dolev_Yao}. Cryptographic primitives are unbreakable but their operation is known by the
adversary, so are all public keys, but private keys are only known by their respective owners.
Moreover, the attacker can obtain any message passing through the network, initiate a communication
with any other entity, and become the receiver of any transmission. The objective of this attacker
is to obtain a certificate for a victim's domain, and then perform an impersonation attack using
that certificate and the corresponding private key. The adversary's goal is to remain undetected as
long as possible if the attack succeeds. \name is designed to completely prevent impersonation
attacks under a set of assumptions, but even if the attacker's capabilities go beyond these
assumptions, attacks can be detected. For this reason, we use two adversary models: one for
prevention and one for detection. Also, because \name supports trust heterogeneity, our adversary
model can only be defined from the perspective of one client establishing a connection to a web
server with specific domain name. Let $N$ be the domain name in question, $f(N)$ the set of CAs
highly trusted by the client for that name, $g(N)$ the set of non-highly trusted CAs for $N$, and
$M$ the set of map servers that the client uses.

\noindent
\textbf{Adversary Model 1 (Prevention):}
The attacker may compromise all CAs in $g(N)$ and a number of map servers such that a
subset of map servers $M_1 \subseteq M$ remains uncompromised and the map servers in $M_1$
collectively support all CAs in $f(N)$. 

\noindent
\textbf{Adversary Model 2 (Detection):}
The attacker may compromise all map servers in $M$ and all CAs in $f(N)$ and $g(N)$.

\section{Overview of \name{}}
\label{sec:overview}

We seek to accomplish two main goals. First, browser vendors and users can define a validation
policy, i.e., label CAs as highly trusted, trusted, and untrusted for each domain to decide which
CAs should be prioritized in case of conflict.  Second, to clearly identify these conflicts, domain
owners must be able to define domain policies. No attacker should then be able to hide or downgrade
these policies. Therefore, it is necessary to provide clients with a comprehensive view of all
certificates, policies, and revocation messages relevant to the domain they are contacting. In the
current ecosystem, this data cannot be easily obtained.

We introduce an entity called \emph{map server}, which provides a comprehensive view of certificates
for its supported set of CAs. The goal of the map server is to aggregate certificate-related data in
a verifiable manner and provide a meaningful interface to both clients and domain owners.  Map
servers use a sparse Merkle hash tree to effectively produce proofs of presence or absence. The data
provided by map servers complements the traditional certificate validation procedure. A user gains a
higher degree of assurance that the binding between a public key and a name is authentic, by
checking that there exists no conflicting certificate for the domain in question.

\begin{figure}[h!]
	\centering
	\includegraphics[width=\columnwidth]{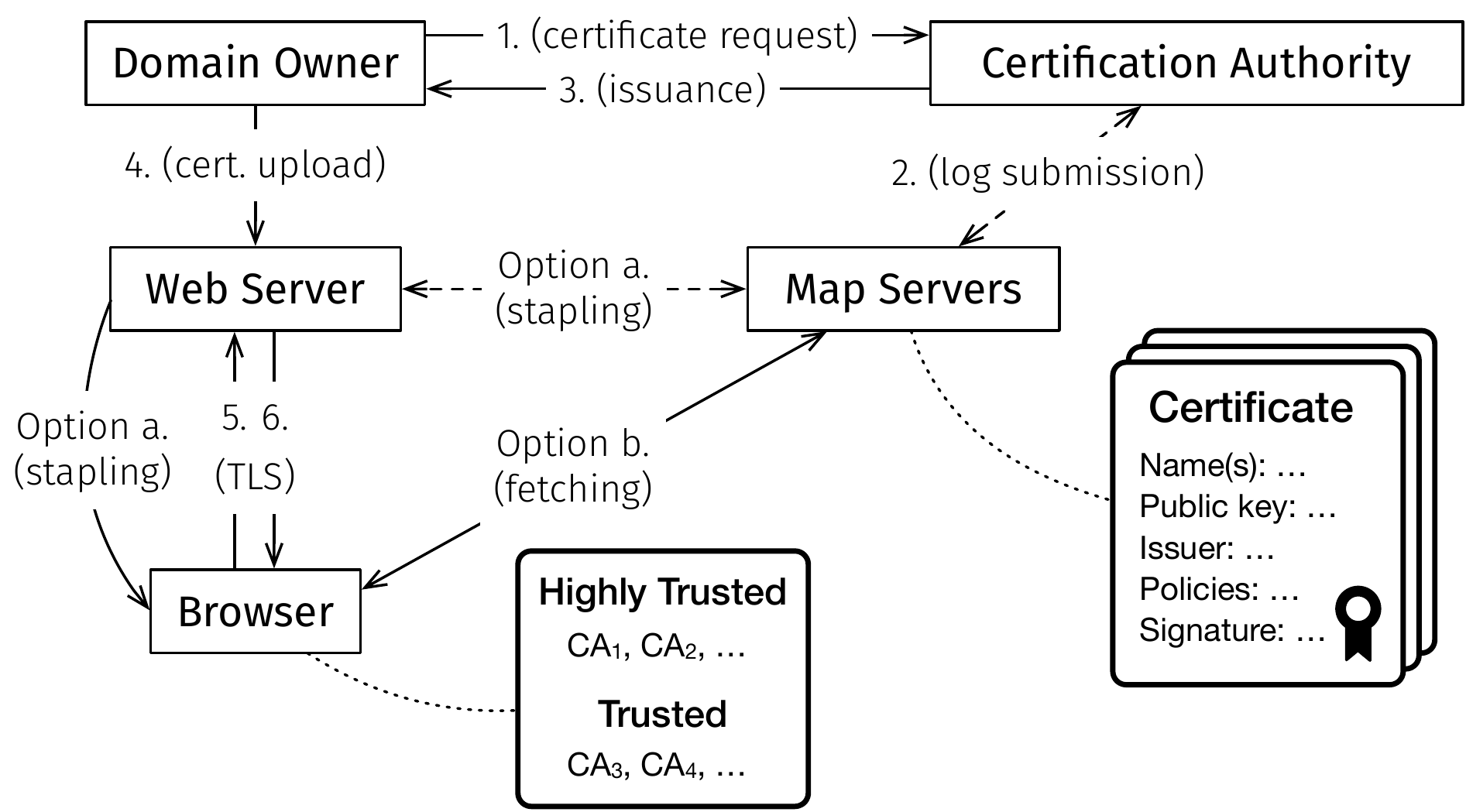}
	\caption{Overview of communication flows before and during the establishment of an HTTPS
          connection. The dashed lines indicate asynchronous communications. \emph{Map servers}
          are introduced in this work and described in Section~\ref{sec:verifiable_logging}.}
	\label{fig:overview}
\end{figure}

Figure~\ref{fig:overview} illustrates how the different entities interact. At a high level, the
steps needed to establish a secure HTTPS connection within \name{} are the following:
\begin{enumerate}[1.]
	\item The domain owner requests a certificate for \emph{www.ex\-am\-ple.com} from a
	certification authority. The certificate signing request sent to the CA may contain additional
	domain policies we introduce in Section~\ref{sec:cert_policies}.
	\item The map server maintains a complete set of certificates (by periodically fetching
          certificates from CT logs, or acting as a special CT log).
	\item The CA returns the certificate to the domain owner. The certificate should contain the
          parameters and policies the domain owner specified in Step 1.
	\item The domain owner configures the web server so that it uses the newly obtained certificate.
	\item The browser connects to the web server via HTTPS\@.
	\item As part of the TLS handshake, the browser receives the certificate, and possibly stapled
	      data periodically fetched by the web server from the map server (see below).
\end{enumerate}

Two main options allow the client to obtain a comprehensive set of certificates for
\emph{www.ex\-am\-ple.com} and the corresponding inclusion proof from a map server:
\begin{enumerate}[a.]
	\item The web server fetches the certificate set and inclusion proof periodically or on-demand
	and provides it to the browser by stapling it to the TLS handshake.
	\item The browser fetches the certificate set and inclusion proof from the map server, directly
	or via DNS (we describe the DNS technique in more detail in \cref{sec:proof-delivery}).
\end{enumerate}

\section{\name{} in Detail}

In this section, we describe in detail the operation of \name{} as well as the data structures upon
which it relies. We start by discussing validation policies that dictate which CAs are more trusted
than others for a given name. Then we present policies that domain owners may define through X.509v3
extensions to opt-in and benefit from the stronger certificate validation that we propose. Then we
describe map servers as an essential component to support the previously defined policies and
certificate revocation. Finally, we discuss how certificates and proofs can be delivered from map
servers to clients in an efficient and privacy-preserving way, and we describe how certificates are
validated.

\subsection{Validation Policies (Trust Levels)}\label{sec:valid-polic-trust}

Clients can specify which CAs they highly trust. We envision that such policies would be defined by
browser vendors, but users should be free to modify their default policies. Validation policies
govern the behavior of $f(N)$ (see \cref{sec:model}), which we defined as the set of authorities
that are highly trusted for name $N$. Below are some examples of policies browser vendors and/or
users may want to define:
\begin{itemize}
	\item[\textbf{CA-Based Policies}:] Some CAs may be more trusted than others, regardless of the
	domain name. This judgment may be based on past events (such as security incidents), the
	validation methods that the CA employs, a reputation for following best practices, and/or
	geopolitical factors.
	As a concrete example of a reason to trust a CA over others, Let's Encrypt now validates domains
	from multiple vantage points or ``perspectives''~\cite{MultiPerspectives} to mitigate routing
	attacks on BGP~\cite{bamboozling,birge2021experiences}.
	\item[\textbf{TLD-Based Policies}:] CAs operating in a given country or region (as defined in
	their certificate) can be designated as more trusted for certain top-level domains (TLDs). For
	example, a policy may state that American CAs are more trusted for domain names ending in
	``.us'' and ``.gov''~\cite{gov_domain}, while Chinese CAs are more trusted for ``.cn''.
    \item[\textbf{Organization Policies}:] Employees might be required to comply with policies for
    domains owned by their company. Online banking or trading platforms, as security-critical
    applications, may also provide their customers with policies they should enforce when connecting
    to their websites, specifying which CAs are highly trusted.
\end{itemize}

We also envision that such policies could be downloaded by users in the form of a ``trust package''
provided by organizations such as the Electronic Frontier Foundation, the CA/Browser Forum, the
Mozilla Foundation, or ICANN.

The fact that different clients could use different validation policies is part of \name{}'s design.
The consequence of this design choice is that---when a conflict appears in the form of a certificate
not respecting a policy signed by a CA---some clients might still be able to establish a connection
to the website (if they do not highly trust the CA that signed the policy), while others will
abandon the connection. In the worst case where a client highly trusts a malicious or compromised
CA, the security of \name{} is equivalent to that of the current web PKI, where a fraudulent
certificate issued by a trusted CA is accepted.

\subsection{Domain Policies}
\label{sec:cert_policies}

Domain owners can also define policies, in this case through X.509v3 extensions. There are two
reasons for this: security and backward compatibility.  Domain policies allow domain owners to
provide stronger validation of their certificates, for example, prohibiting wildcard certificates or
listing authorized issuers. Relying parties receive all relevant certificates through map servers so
an attacker with a fraudulent certificate cannot hide any policy.  Domain policies also act as an
opt-in feature, which is required since \name{} mandates a stronger certificate validation procedure
that could potentially break validation of existing certificates.  Specifically, a certificate may
contain the following attributes:
\begin{itemize}
	\item[\texttt{ISSUERS} (\textit{Set\_Attribute}):] set of public keys (CAs) that may be used to
	verify signatures on this domain's certificates. If this extension is not present, then all the
	CAs the client trusts are authorized to issue certificates.
	\item[\texttt{SUBDOMAINS} (\textit{Set\_Attribute}):] set of subdomain names for which
	certificates can be issued. Ranges of subdomains can be covered with wildcards (e.g.,
	\textit{*.sub.example.com}). If this extension is not present, then all subdomain names are
	authorized.
	\item[\texttt{WILDCARD\_FORBIDDEN} \textit{(Bool\_Attribute)}:] prohibits the use of wildcard
	certificates for this domain.
	\item[\texttt{MAX\_LIFETIME} (\textit{Max\_Attribute}):] max. certificate lifetime.
\end{itemize}

Each attribute is marked as either \emph{inherited} or \emph{non-inherited}. An inherited attribute
will be passed on to subdomains. Non-defined, non-inherited attributes default to the browser
policy.

Domain owners should use a consistent set of policies and certificates. In other words, domain
owners should make sure that they do not generate themselves any policy violations, in order to
guarantee that their website remains available to all clients (i.e., regardless of which CA(s) the
clients highly trust). Policy violations should only be observed when a fraudulent certificate is
issued by a malicious or compromised CA. To let domain owners change their policies over time
without disrupting legitimate connections to their website, domain policies can be revoked through
map servers (certificate and policy revocation are discussed in \cref{sec:verifiable_logging}).

\paragraph{Multiple certificates for the same domain.}
Although we presented \name{} as an alternative to multi-signed certificates (to increase resilience
against CA compromise), the two approaches can be combined. A domain owner can obtain certificates
from several CAs, as long as the certificates respect the domain policies. Ideally, policies and
other parameters defined in the certificates would be identical, but this is impossible to guarantee
if the certificates are issued independently. Therefore, clients must consider the strictest of all
policies.

\paragraph{Policy Resolution.}
Clients resolve domain policies using their default browser policy and a set of X.509v3 policy
extensions.  The strictest policy is calculated by iteratively applying a fold operation on each
attribute of the policy.  \textit{Bool\_Attributes} are combined through logical conjunction
($B_0\wedge B_1$), \textit{Max\_Attributes} are combined by taking the minimum value
($\text{min}(M_0,M_1)$), and \textit{Set\_Attributes} are combined through intersection ($S_0\cap
S_1$).

\subsection{Verifiable Data Structures in Map Servers}\label{sec:verifiable_logging}
\begin{figure*}
	\centering
	\begin{tikzpicture}[scale=.49,every tree node/.style={font=\LARGE,anchor=base}]
	\Tree
	[.\node(root){\textbf{signed map head}};
		[.{hash}
			[.{hash}
				[.{...}
					[.{default} [.{empty} {index: $0$} ] ]
					[.{default} [.{empty} {index: $1$} ] ]
				]
				[.{...}
					[.{...} ]
					[.{$\text{H}(e_1)$} [.{$e_1=\text{Entry(example.\underline{com})}$}
					{$\text{index: H(example.\underline{com})}$} ] ]
				]
			]
			[.{default} {...}
			]
		]
		[.{hash}
			[.{hash}
				[.{...}
					[.{default} {...} ]
					[.{$\text{H}(e_2)$} [.\node(e2){$e_2=\text{Entry(u-tokyo.\underline{ac.jp})}$};
					{$\text{index: H(u-tokyo.\underline{ac.jp})}$} ] ]
				]
				[.{...} ] 
			]
			[.{hash}
				[.{...} ]
				[.{...}
					[.{default} {...} ]
					[.{default} [.{empty} {index: $2^{256}-1$} ] ] ] 
			]
		]
	]
	\begin{scope}[shift={(18,0cm)}]
		\Tree 
		[.{\textbf{signed consistency head}}
	        [.{hash}
				[.{hash}
					[.\node(leaf){SMH$_0$}; ] ]
				[.{hash}
					[.{SMH$_1$} ] ]	
	        ]
	        [.{hash}
	        	[.{SMH$_2$} ]
	        ]
        ]
	\end{scope}
	\begin{scope}[shift={(18,-6cm)}]
		\Tree 
		[.\node(sdth){\textbf{subdomain tree head}};
	        [.{hash}
				[.{...} ]
				[.{...} ]
	        ]
	        [.{hash}
	        	[.{...} ]
	        	[.{...} ]
	        ]
        ]
	\end{scope}
	\draw[->, thin, dashed](root) .. controls +(north east:9) and +(south west:4) .. (leaf);
	\draw[<-, thin, dashed](e2) .. controls +(345:7) and +(195:7) .. (sdth);
	\draw [decorate,decoration={brace,amplitude=5pt,mirror}] (-15.5,-7.4) -- (14.5,-7.4)
	  node[midway,yshift=-3ex]{$2^{256}$ leaves (most are empty)};
	\end{tikzpicture}
	\caption{Sparse Merkle tree with the corresponding consistency tree and a subdomain tree.
	Every leaf in the sparse Merkle hash tree corresponds
	to an entry in a map. Every effective second-level domain has a distinct entry in the main tree.
 	The index of each entry in the tree is determined using the hash of the domain name. This
 	construction can  be used in practice because building the entire tree is not necessary to
 	generate proofs: most nodes have default values.}
	\label{fig:sparse_mht}
\end{figure*}
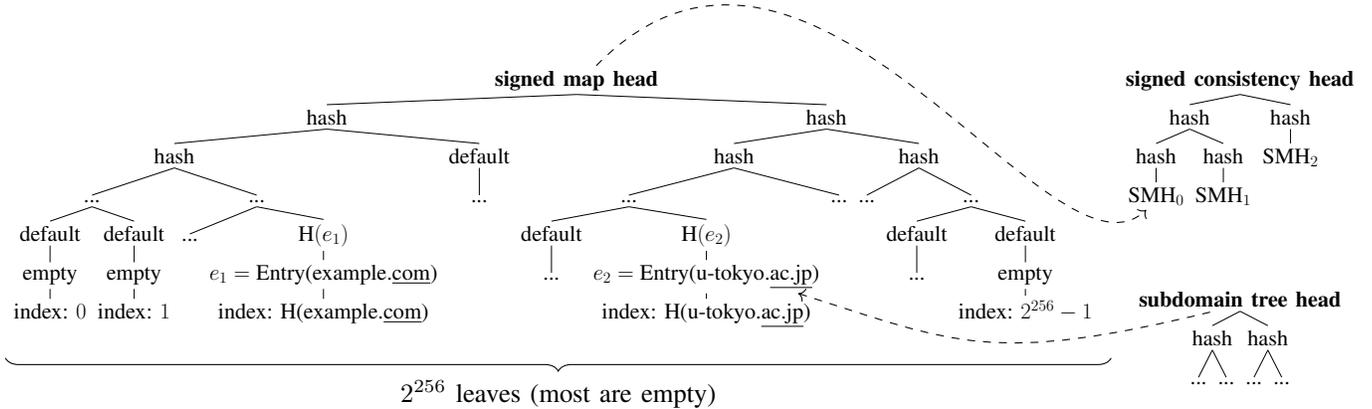

We introduce an entity called map server that provides a mapping from domain names to a set of
certificates and revocations for this domain and parent domains. Each map server supports a set of
CAs and keeps track of certificates issued by these CAs by leveraging the existing CT infrastructure
and log data. The efficient audit of map servers is enabled by sparse Merkle hash trees, which we
extend with nested trees to support hierarchical naming (similar to DNS). To the best of our
knowledge, this is the first proposal for a verifiable data structure that provides a complete view
of all certificate-related data of a domain and its parent domains while supporting efficient proofs
of presence and absence of such data.  Figure~\ref{fig:sparse_mht} gives an overview of the data
structures used by a map server.

\paragraph{Comparison to CT logs.}
Even with Certificate Transparency, it is not possible to query a log server to directly verify that
no certificate has been illegitimately issued for a given domain name. Instead, domain owners must
rely on monitors, which keep entire copies of several logs. Only a few monitors exist at the
moment. Li~et~al.~\cite{li2019certificate} recently reported that none of the active third-party
monitors they found could guarantee to return a complete set of certificates.  The critical
interface missing from log servers at the moment is thus one for fetching all valid (i.e.,
unexpired) certificates for a given domain name. It would seem natural to extend existing CT log
servers to support this operation, but unreasonable to expect all log servers to be
updated simultaneously.

\paragraph{Hierarchical Naming.}
Map servers distinguish between effective second-level domains (e2LD)~\cite{roberts2019you} (i.e.,
domains where the parent domain is a public suffix~\cite{PublicSuffixList}) and descendants of e2LDs
(hereinafter subdomains). The remaining domains---parent domains of e2LDs and domains without a
valid TLD---are invalid and the map server rejects certificates for such domains (e.g.,
\textit{ac.jp} or \textit{test.invalid}). e2LD entries are stored in a single sparse MHT\@. An
advantage of such a hierarchy over a label-based hierarchy such as DNSSEC is that certificates for
e2LDs can be stored at a lower depth (e.g., example.blogspot.co.uk has depth 0 instead of 3 in
DNSSEC), which reduces the proof size.
The entries of subdomains are stored in nested sparse MHTs located below the parent domain's entry.
The data structure for subdomains is the same as for the e2LDs, except that the key used for the
index calculation only includes the name of the subdomain (without the parent domain). Since the
Merkle proof of a subdomain contains all parent-domain entries, policies issued by
parent domain owners are included in the subdomain proof. A parent domain owner can thus restrict
CAs from issuing certificates for its subdomains or allow certificates only for certain subdomains.
The reason map servers reject entries of public suffixes is that a certificate for a public suffix
would enforce policies specified in the certificate for all e2LD domains using this public suffix.
An algorithm for constructing the map server as described above is presented in
\cref{sec:map_server_construction}. The map server is also pruned periodically to remove expired
certificates.

\paragraph{Entry Format.}
The map server creates an entry for each domain with at least one certificate (valid or revoked) or
one active subdomain. The entry consists of the certificates and revocation messages of both the
domain (\textit{example.com}) and the corresponding wildcard (\textit{*.example.com}), and the
root of the subdomain MHT, as shown in~\cref{tab:entry}. Each certificate might contain policies in
the form of X.509v3 extensions. The wire format of an entry is its ASN.1 DER
encoding~\cite{itu690.2015} where certificates, revocation messages, and tree roots are encoded as
octet strings or sequences of octet strings.

\begin{table}[t]
  \centering
  \small
  \begin{tabular}{lll}
    \toprule
    Content & Type & Domain(s) \\
    \midrule
    Certificates & list$<$X.509 certificate$>$ & \textit{example.com} \\
    Revocations & list$<$revocation message$>$ & \textit{example.com} \\
    Certificates & list$<$X.509 certificate$>$ & \textit{*.example.com} \\
    Revocations & list$<$revocation message$>$ & \textit{*.example.com} \\
    Tree Root & cryptographic hash & [subdomains] \\
    \bottomrule
  \end{tabular}
  \caption{The map entry for \textit{example.com}.}\label{tab:entry}
\end{table}

\paragraph{Revocations.}
An end-entity certificate can be revoked either by a CA in the certification path (using the private
key that corresponds to the CA certificate) or by the domain owner (using the private key that
corresponds to the certificate itself). A revocation message for certificate $C$, revoked using a
private key $k$, has the form $R_{C}=\text{Sig}_{k}(H(C),\text{revoke})$. Such a certificate
revocation message can either be pushed to map servers by the domain owner or sent to the issuing CA
which forwards the message to map servers. A policy can be nullified by revoking the certificate in
which it was defined. It is possible to revoke a policy but not the certificate in which it is
defined. This option will help domain owners change their domain policies over time without
disrupting legitimate connections that still rely on a previous certificate. It will also help them
recover from erroneously defined policies.

\paragraph{Sparse MHT as Key-Value Store.}
A sparse Merkle tree, as shown in \cref{fig:sparse_mht}, provides a mapping from keys (domains) to
values (entries). Each leaf corresponds to a key-value pair. The leaf's position is determined by
the hash of the key, leading to $2^{256}$ possible leaves for a hash output length of 256 bits. The
hash of a leaf is the hash of the concatenation of a fixed leaf prefix (e.g., 0x00) and the DER
representation of the corresponding entry. The hash of an intermediate node is the hash of the
concatenation of a fixed node prefix (e.g., 0x01) and the hash of the left and right child.

The path from the root to a leaf is defined as follows: At level $i$, if the $i$th bit of the hash
is 0, the left child is selected, otherwise the right child is selected. Hence paths have a fixed
length of 256 and the proof that a certain key with a given value exists consists of the 256 hashes
of siblings that are necessary to construct the hash chain to the signed tree root.

The non-existence
of a key is proven by showing that the value at the position of the hash of the key is the default
value (indicating an empty leaf) which is distinct from any real value. Such a proof is large
($256\cdot{}256$ bits = \SI{8}{\kilo\byte}) but can be compressed since the majority of the leaves
in a sparse tree are empty. The idea of the compression is that hash values for subtrees where all
leaves are empty can independently be (pre-)computed if the default value and the level are known.
The map server then omits all values that can independently be computed when transmitting a proof.
The sparse MHT in \cref{fig:sparse_mht} with the public suffixes \textit{com} and \textit{ac.jp}
stores entries for the e2LDs \textit{example.com} and \textit{u-tokyo.ac.jp}.

\paragraph{Reaping the Benefits of Sparse MHTs.}
Since each e2LD has a separate tree for its subdomains, adding subdomains only increases
the size of the given subtree and thus does not affect proof sizes of other domains.  In a regular
MHT, proofs of presence have a size of $\ceil*{\text{log}_2L}$ where $L$ is the number of leaves
(unless otherwise stated, we express proof sizes in number of hash values required to verify the
proof). A proof of absence, however, is up to twice as large as it requires inclusion proofs of the
previous and next leaf entry. In sparse MHTs, proofs of presence and absence have the same size,
since the absence of a domain can be verified with a single inclusion proof. In a sparse MHT, the
proof size depends on the structure of the sparse MHT\@. Only non-default nodes are considered part
of the proof since default nodes can be pre-calculated and stored. If the non-empty leaves are
uniformly distributed, the average proof size is minimized and the same as in a regular MHT
$\ceil*{\text{log}_2L}$. In sparse MHTs, adding, updating, and removing entries require changing
$\ceil*{\text{log}_2L}$ node values on average since each on-path node must be updated.

\paragraph{Proving Consistency over Time.}
In addition to the sparse e2LD MHT, the map server maintains a consistency MHT with chronologically
ordered signed map heads (SMH) coming from the sparse e2LD MHT, as shown in~\cref{fig:sparse_mht}.
The consistency tree can prove that the modifications between log revisions are correct, and prove
that the log did not include non-existing certificates or exclude existing certificates. In other
words, the signed map head represents the map at a given point in time, while the signed consistency
head represents the entire history of the map.

\subsection{Selection of Map Servers}
\label{sec:server-selection}

Each client must have at least a set of map servers such that each highly trusted CA is supported by
a quorum of servers in this set. Typically, each map server would support all major CAs and the
quorum would be equal to 1 (especially during the initial deployment of \name{}). Therefore, clients
would only need proofs from a small set of map servers. It is recommended, however, that the set of
map servers be slightly larger than what is required by the quorum to permit failover in case a map
server is unreachable. When different map servers are contacted, if they give different replies, the
client considers all policies signed by highly trusted CAs, as specified in \cref{alg:validation}.

For validation policies where a single set of CAs is highly trusted for all domain names, only one
set of map servers is needed. For more complex validation policies, a natural way to define those
policies as well as the map servers to use in each case is to use a list of tuples of the following
form:
\begin{align*}
	&\langle N_1, f(N_1), M_1 \rangle, \\
	&\langle N_2, f(N_2), M_2 \rangle, \\
	&\dots
\end{align*}
where $N_1$ is a set of domain names, $f(N_1)$ is the set of highly trusted CAs for $N_1$, $M_1$ is
the set of map servers to use for $N_1$, and $M_1 \cup M_2 \cup \dots = M$ (see
\cref{sec:adversary_model}). 

An algorithm to select an appropriate set of servers automatically is presented in
\cref{sec:map_server_selection_algo}. The algorithm performs the selection by solving a set
multicover problem using a greedy approach. Using this algorithm with the set of all map servers
$M$, the user can change validation policies without having to manually select the set of map
servers to use.

Although \name{} is designed to be flexible and expressive, we expect this process to be
straightforward, based on a fairly low number of map servers and handled by browser (or OS) vendors
rather than end users in most cases (i.e., unless the user wishes to change their default settings).

\subsection{Proof Delivery}\label{sec:proof-delivery}
 
\paragraph{Stapling.}
The first approach for delivering proofs to clients is to have the web server embed
them into a TLS extension, similarly to how revocation messages can be delivered with OCSP stapling.
This technique does not cause privacy issues but requires that servers be updated to
fetch proofs related to their own name periodically from a set of map servers.

\paragraph{Fetching via DNS.}
In case the web server does not embed relevant proofs from a map server in its response, the client
should fetch such proofs on its own.
Instead of directly contacting map servers, the client can fetch log data through
DNS~\cite{CT_over_DNS,szalachowski2017short}.
For example, to request a proof from a map server \emph{mapserver1.net} for the domain name
\emph{www.example.com}, the client can send a recursive DNS query for
\emph{www.example.com.map\-server\-1.net} to its DNS resolver. The DNS resolver finds the IP address
of the map server and forwards the client's query to that address. Acting as a name server, the map
server replies with one or more TXT records\footnotemark\xspace containing the entry
of \emph{www.example.com}. Finally, the resolver returns the records to the client. \footnotetext{We
employ TXT records in this design, as well as in our proof-of-concept implementation, in order for
\name{} to be compatible with existing infrastructure, and leave the introduction of a
standardized resource record type dedicated to our purposes as potential future work.}
The advantages of using DNS for the purpose of delivering proofs from map servers to clients are
numerous:
\begin{itemize}
	\item[\textbf{Decentralization:}] The decentralized and hierarchical nature of DNS falls in line
	with our objective of avoiding reliance on a global authority.
	\item[\textbf{Caching:}] Records can be cached at several levels, for a configurable amount of
	time (TTL), to minimize latency.
	\item[\textbf{Privacy:}] Although DNS is not in itself a privacy-preserving system,
	the resolver is always aware of the domain names the client is trying to reach, so the privacy
	implications of asking a public DNS resolver for extra information about those names are
    limited.
	\item[\textbf{Timeliness:}] DNS resolution occurs before the TLS connection is established
	(to obtain the web server's IP address), which minimizes perceived page-load time increase.
\end{itemize}

DNS already supports the delivery of certificate-related data (e.g., through CAA, CERT, and TLSA
records). DANE~\cite{rfc7671}, in particular, binds certificates to domain names directly in the
name system with the objective of solving the issues caused by rogue CAs. Unfortunately, DANE relies
upon DNSSEC for protecting name-to-certificate bindings, and is not widely supported by browsers at
the moment. In contrast, \name{} does not require an authenticated name system such as DNSSEC,
because the cryptographic objects we transport through DNS are authenticated independently, i.e.,
users know the public keys of their trusted map servers.

One concern over using DNS for transporting cryptographic data is that the size of messages sent
over UDP is limited by the DNS standard. However, RFC 6891~\cite{rfc6891} describes extension
mechanisms to enable using UDP for messages with sizes beyond the limits of traditional DNS. The RFC
suggests that requestors try initially selecting a maximum payload size of 4096 bytes, which is
sufficient to contain a cryptographic proof (such as the ones produced by Trillian) in many cases.
If necessary, falling back to TCP enables the transport of larger packet sizes, at the price of
increased latency.

\paragraph{Alternative Delivery Methods.}
Directly fetching a proof about a specific domain from a map server has privacy implications, but
there might be situations where the user is willing to send requests directly to a map
server. In particular, if the user relies on a public DNS server (such as CloudFlare's 1.1.1.1 or
Google's 8.8.8.8), and if one of the company operating the name server also operates map servers,
then the privacy implications of fetching proofs directly are limited (assuming a secure channel is
established between the client and the map server). Another approach would be to use a middlebox
(instead of updating the web server) to staple proofs by having the middlebox detect new
connections and append relevant data to the TLS handshake (which is in plain
text)~\cite{RITM2016,lee2019matls}.

\subsection{Certificate Validation}
\label{sec:validation}

The certificate provided by the web server during the TLS handshake is validated by the client using
the normal validation procedure and additional data provided by the map server(s), which may include
a list of other certificates for the same domain, a list of certificates for parent domains, and a
list of revocations for both. The client will then check revocations, resolve the domain's policy
(considering only policies defined by highly trusted CAs), and verify that the certificate respects
the resolved domain policy.

\begin{algorithm}[h]
	\small
	\DontPrintSemicolon
	\SetKwIF{If}{ElseIf}{Else}{if}{}{else if}{else}{end if}
	\SetKwProg{function}{function}{}{}
	$n:$ domain name for which the certificate is validated \\
	$f(n):$ CAs highly trusted for $n$ (see Section~\ref{sec:model}) \\
	$C:$ certificate received during TLS connection \\
	$C_{\textit{list}}:$ list of certificates for $n$ and parent domains \\
	$R:$ dictionary (key: certificate hash; value: revocations) \\
	\Comment{$C_{\textit{list}}$ and $R$ come from map server(s)} \\
	$P_{\textit{browser}}$: default browser policy \\
	\function{\textsc{Validate}($n$, $f$, $C$, $C_{\textit{list}}$, $R$)}{
		\If{\normalfont \textbf{not} legacyValid($C$)}{
			\Return false \Comment{legacy validation failed}
		}
		\For{\normalfont $r \in R[\text{hash}(C)]$}{
			\If{$r$.isValid()}{
				\Return false \Comment{certificate is revoked}
			}
		}
		\For(\Comment{filter certificates}){$c \in C_{\textit{list}}$}{
			\If{\normalfont \textbf{not} (legacyValid($c$) \textbf{and} signedBy($c$, f($n$)))}{
				remove $c$ from $C_{\textit{list}}$; continue
			}
			\For(\Comment{check revocations}){\normalfont $r \in R[\text{hash}(c)]$}{
				\If{$r$.isValid()}{
					remove $c$ from $C_{\textit{list}}$; break
				}
			}
		}
		$p \gets P_{\textit{browser}}$ \Comment{start with default browser policy} \\
		\For{$c \in C \cup C_{\textit{list}}$}{
			$p' \gets c.\textit{policy}()$ \Comment{embedded policy} \\
			\For{$a \in \textit{All\_Attributes}$}{
				\If{$p'[a]\text{.isInherited}()$ \textbf{or} $n \in c.\text{names}()$}{
					\If{$a \in \textit{Bool\_Attributes}$}{
						$p[a] \gets (p[a]\;\texttt{\&\&}\;p'[a])$
					}
					\If{$a \in \textit{Max\_Attributes}$}{
						$p[a] \gets \min(p[a], p'[a])$
					}
					\If{$a \in \textit{Set\_Attributes}$}{
						$p[a] \gets (p[a] \cap p'[a])$
					}
				}
			}
		}
		\If(\Comment{issuers, wildcard, ...}){\normalfont ViolatesPolicy($C$, $p$)}{
				\Return false
		}
		\Return true
	}
	\caption{Certificate Validation}
	\label{alg:validation}
\end{algorithm}

Algorithm~\ref{alg:validation} shows the certificate validation logic of \name{} in pseudocode. The
validation function takes six inputs: the domain name of the web server,
the function $f$ representing CAs highly trusted for that name,
the certificate received during the TLS handshake, other certificates for the same domain that the
map server may have returned in the previous step, a set of revocations for the above certificates,
and the browser's default policy. The algorithm starts by verifying that the legacy X.509 validation
procedure succeeds and that the certificate is not revoked. Then, invalid and revoked certificates
are removed from the list of additional certificates. Certificates issued by non-highly trusted CAs
are also removed from the list. We then iterate through these certificates and resolve the domain's
policy, considering the strictest of all policies in the received certificate and the filtered list
of additional certificates. Finally, the algorithm checks that all domain policies are respected.

\section{Security Analysis}

\name{} provides strictly stronger security guarantees than today's web PKI since any certificate
rejected by a browser today would also be rejected by \name{}. The simplicity of our certificate
validation algorithm (Algorithm~\ref{alg:validation}) allows a direct demonstration of this
property. The first step of that algorithm is to run the legacy certificate validation procedure; if
it fails, the algorithm returns immediately. This is true even if all map servers are malicious and
colluding. We now discuss the additional security guarantees of \name{} and potential attack
vectors.

In the analysis below, we consider the perspective of a user with
browser $B$, which uses a set of map servers that do not necessarily support all CAs but
periodically fetch the data from all the CT log servers supported by browser $B$.

\subsection{Attack Prevention}

Recall that we presented two adversary models in Section~\ref{sec:adversary_model}: one for attack
prevention and one for attack detection. We start by analyzing the security guarantees that \name{}
provides to a client under the assumption that highly trusted CAs and a subset of map servers are
honest and behaving correctly (i.e., Adversary Model 1).

\emph{If the client considers $\textit{CA}_1$ highly trusted for name $N$, and $\textit{CA}_1$ has
issued a certificate $C_1$ (which is valid) for $N$, then the client will not accept any certificate
that does not comply with the policies in $C_1$.}

The above statement holds because, by assumptions of Adversary Model 1, highly trusted CAs are
honest and at least one map server $M_1$ in the set of map servers that the clients uses supports
$\textit{CA}_1$. Therefore, the attacker can obtain a valid certificate from a non-highly trusted CA
but cannot hide $C_1$ from the client because $M_1$ (which is honest by assumption) will show the
certificate to the client. Also, the attacker cannot define illegitimate policies for $N$ because
only the policies defined by highly trusted are considered by the client.

\subsection{Attack Detection (CA Misbehavior)}

In Adversary Model 2, we relaxed our assumptions to include the possibility that all CAs and all map
servers are malicious. The illegitimate issuance of a certificate by a highly trusted CA cannot be
prevented, but it can be detected.

\emph{If a client considers $\textit{CA}_1$ highly trusted for name $N$, and $\textit{CA}_1$ has
illegitimately issued a certificate for $N$, which is used in an attack, then the domain owner
will be able to detect the illegitimate certificate.}

CT already allows detecting CA misbehavior. Given that \name{} is designed to complement the current
web PKI, the detection of such attacks is naturally possible. But map
servers facilitate the detection of CA misbehavior as they provide a more powerful API than CT:
all the certificates related to a given domain can be obtained with a single query, whereas CT
relies on external auditors that keep entire log copies to detect fraudulent certificates.
These auditors must in turn be trusted to perform this task correctly and, unlike map servers, do
not use verifiable data structures.
The enhanced transparency that \name provides can then be leveraged by clients and browser vendors
to select their highly trusted CAs.

\subsection{Dealing with Malicious Log Servers}

Submitting every new certificate to several CT logs is already a requirement. Chrome, for example,
requires that each certificate come with evidence that it was submitted to at least one
Google-operated log and one non-Google-operated log~\cite{chrome_log_policy}. By ensuring that
certificates are logged by a large and diverse set of log servers, misbehavior can be tolerated.
Only if several log servers are compromised and colluding can they hide the existence of
certificates and policies from map servers.

Additionally, violation of the append-only property can be efficiently detected by external auditors
using the MHT's properties. Auditors can also verify that log servers include certificates after the
maximum merge delay (MMD). \name{} favors log servers with small MMDs since this allows the map
servers to have an up-to-date view of the certificate ecosystem.
Furthermore, a gossip protocol~\cite{gossip2015,ietf-trans-gossip-05} can be used to detect a
split-view attack in which a log server would consistently provide different views of its MHT to
different clients.

\subsection{Dealing with Malicious Map Servers}

Each map server returns a single proof for a given (sub)domain including the certificates and
possibly revocations of all parent domains. If the user-selected set of trusted map servers provide
a complete set of certificates and revocation messages, the user can use the provided domain
policies to verify the validity of a certificate. A malicious map server, however, can hide
certificates and revocation messages by not including them into its MHTs or providing different
views to clients.

A malicious map server could circumvent policies by removing the corresponding certificates. The map
server could, for example, remove a parent domain policy that restricts subdomains for which
certificates can be issued. However, this attack is only useful if the map server colludes with a
CA.

There are two mitigations to reduce the impact of malicious map servers. The first mitigation is the
client's ability to specify which map servers are trusted and how many trusted map servers need to
support each highly trusted CA (quorum). This protects against attackers that compromise up to
$\text{quorum} - 1$ selected map servers. The second mitigation is, as for CT log servers, that the
audit of map servers is facilitated by Merkle hash trees. External auditors can verify the
correctness of a map server's operation, such as the append-only property, completeness with regard
to supported log servers, and update interval. As a last line of defense, a gossip protocol can also
be used to verify the consistency of map servers.

\subsection{False Positives}
Some certificates that would be valid today might be rejected in \name{}. This stronger certificate
validation may raise concerns over false positives. We stress, however, that domain owners must
opt-in (by obtaining a certificate with appropriate certificate extensions from a CA) to reap the
benefits of \name{}. A domain owner may forgo strict domain policies to prioritize availability over
security. Moreover, Algorithm~\ref{alg:validation} states that only policies signed by highly
trusted CAs are considered (the other certificates are filtered before policy resolution).
Therefore, an attacker who compromises a CA that is not highly trusted by the victim cannot falsely
define a policy. An attacker who manages to obtain a certificate from a CA that is highly trusted by
some clients may cause a legitimate certificate to be rejected by those clients, but this will
become evident through logging. The incriminated certificate will then be revoked and the CA in
question will suffer the consequences of the security breach. \name{} thus prioritizes security over
availability in case a highly trusted CA is compromised.

\subsection{Denial-of-Service Vectors}

Some attacks do not threaten our main security guarantees but could impede the availability of the
system. For example, an attacker can launch a resource-exhaustion attack by injecting certificates
for e2LDs such that the hash of the injected domain and target domain share the same prefix. This
would increase the proof size of the target domain in the sparse MHT\@.  The impact of such attacks
is very low due to the complexity of finding alternative inputs, whose hash partially matches the
attacked domain's hash.  Mitigations for these attacks are including an unpredictable, regularly
updated value in the sparse MHT's hash calculation or using verifiable data structures with fixed
proof sizes, such as sorted-list MHTs. Denial-of-Service attacks and their mitigations are explained
in more detail in~\cref{sec:denial-of-service-vectors-and-mitigations}.

\section{Realization in Practice}
\label{sec:impl}

We implemented a map server prototype that manages sparse MHTs using Trillian~\cite{Trillian} and
distributes proofs to clients via DNS using CoreDNS~\cite{CoreDNS}, a DNS server implementation that
supports plugins. Additionally, we implemented a browser extension that verifies certificates used
in the TLS handshake of every website and blocks connections using non-policy compliant
certificates. The web extension is implemented using the Mozilla WebExtensions
API~\cite{MozillaWebExtensions}, a cross-platform JavaScript API for browser extensions.  A concrete
implementation would staple proofs using a TLS extension analogous to OCSP stapling.  For our
performance evaluation, we simulate stapling by encoding map server proofs within X.509 extensions.
We will make our code base publicly available. In this section, we present two concrete use cases
based on domain validation procedures and organization policies. We also evaluate the overhead of
our system and the performance of our proof-of-concept implementation.  To do so, we created a map
server, added over one million certificates from a Google operated CT log (Argon 2019), and
evaluated the number of (unique) certificates, proof sizes, and processing times on the map server.
Finally, we discuss deployment of map servers in the real world.

\subsection{Use Case 1: Multi-Perspective Validation}\label{sec:use-case-multi}

We first present a use case that demonstrates how \name{} can enable security innovation in the CA
ecosystem. In particular, an attacker must not be able to downgrade a certificate issued by a
highly-trusted CA to a certificate issued by a non-highly trusted CA\@. In today's ecosystem, CAs
that aim to innovate via stronger methods for domain validation are limited in their ability to
provide security benefits to clients and domain owners, because any other vulnerable CA can issue
certificates for the same domains.

We demonstrate that our prototype can prevent attacks and favor non-vulnerable CAs, with minimal
software and operational changes. We consider a scenario where the attacker wants to perform a
man-in-the-middle attack and obtains a bogus certificate from a regular CA through a BGP hijacking
attack during the domain validation to trick the CA into issuing the certificate~\cite{bamboozling}.
We simulate this attack by creating a CA certificate, labeling it as non-highly trusted, and signing
the bogus certificate with our newly created CA\@. In order to protect against such attacks,
``multi-perspective domain validation'' prevents attackers that do not have the capability to
perform BGP hijacking attacks on all vantage points simultaneously from obtaining an illegitimate
certificate. Let's Encrypt, for example, has been supporting multi-perspective domain validation
since February 2020~\cite{MultiPerspectives,birge2021experiences}.

We obtained a certificate from Let's Encrypt for the domain under attack. Because Let's Encrypt does
not support custom certificate extensions, we defined as our default browser policy the
\texttt{ISSUERS} attribute to only contain the public keys of Let's Encrypt root CA certificates.
Our legitimate certificate was automatically submitted to CT logs by Let's Encrypt and appended to
the corresponding trees within a day. We defined Let's Encrypt as a highly trusted CA, and made sure
the two certificates were added to our map server. We then installed our browser extension, added
our prototype map server as a trusted map server, and evaluated the man-in-the-middle attack. Before
connecting to the attacker's website, our browser extension now sees in the reply from the map
server that the legitimate certificate is signed by a highly trusted CA (Let's Encrypt) and does not
allow any certificate with a different public key to be considered valid. Our browser extension thus
blocks the connection to the attacker's server.

Multi-perspective validation is just one example of a reason to trust a CA over others. The
assessment of a CA's trustworthiness~\cite{heinl2019mercat,kumar2018tracking} can based on many
other criteria, such as compliance to IETF standards~\cite{rfc5280} and CA/Browser Forum
requirements~\cite{cab_baseline_requirements}.

\subsection{Use Case 2: Organization Policies}

Another situation in which \name would prove particularly beneficial is the following. Consider an
organization with high security requirements (e.g., a government agency) operating numerous websites
and web services. Assume that this organization relies on a single CA to obtain its certificates.
The employees of that organization (and other clients relying on it) could take advantage of \name{}
to avoid falling victim to a MITM attack launched by a foreign state that controls a CA, for
example.

On the server side, this only requires defining an \texttt{ISSUERS} policy that contains the CA in
question (through a certificate extension). On the client side, there are multiple ways to benefit
from \name{} in this situation. The most obvious is for clients to use a browser that supports
\name{} and make sure that the CA in question is highly trusted (or install a ``trust
package'' in which the CA is marked as highly trusted). But there are other options that
clients could use if \name{} is not supported by major browsers. First, the organization can
develop a browser plugin. Second, some organizations recommend using their own custom ``secure
browser''~\cite{Soliton,Coronic}; \name{} could be built into such a browser with the appropriate
validation policies. Finally, if the organization has an app that uses an API over HTTPS, then the
app can be configured to support \name{} for a more secure communication with the API.

\subsection{Performance Evaluation}

We now evaluate four performance metrics: (a) latency, (b) proof size, (c) proof generation time,
and (d) typical number of subdomains per domain. We used three certificate datasets for this
evaluation:
\begin{itemize}
	\item \textbf{Alexa Top 1K:} We scraped $\sim$700 certificates from the 1000 websites of the
	Alexa top 1K list. We only used this dataset in our latency evaluation to get a relatively
	small set of commonly used certificates that our test client could sequentially visit.
	\item \textbf{Single CT Log:} We fetched $\sim$1 million certificates from a single log server
	(Google Argon 2019). We used this dataset to analyze the effects of increasing the total number
	of certificates on proof generation (time and size), and to evaluate the scalability of our
	proof-of-concept implementation of a map server.
	\item \textbf{All Google-Operated Logs:} We fetched $\sim$20 million certificates from all
	Google-operated log server. We used this dataset to infer the typical number of subdomains of
	each domain.
\end{itemize}

\paragraph{Latency.}
The metric of interest for us here is the time to first byte (TTFB), which is the duration from the
moment when the browser initiates a page request (including the DNS lookup) to the moment when the
browser starts receiving HTTP data. We define the TTFB overhead as the additional time it takes
before receiving the first byte when using \name{} (compared to a regular HTTPS request). Our
evaluation shows that the DNS and stapling approaches typically have a TTFB overhead of less than
\SI{1}{ms} and \SI{10}{ms}, respectively. For this evaluation, we set up a client in our university
network and server on Digital Ocean~\cite{digitalocean}. The server scraped the TLS certificates
from the 1000 most popular domains that enabled HTTPS (resulting in $\sim$700 certificates) and
added the certificates to a local map server. The client installed the browser plugin and
sequentially visited these domains.

\begin{figure}[t]
  \centering
  \begin{subfigure}{\columnwidth}
	\centering
	\vspace{2ex}
  	\includegraphics[width=.8\linewidth]{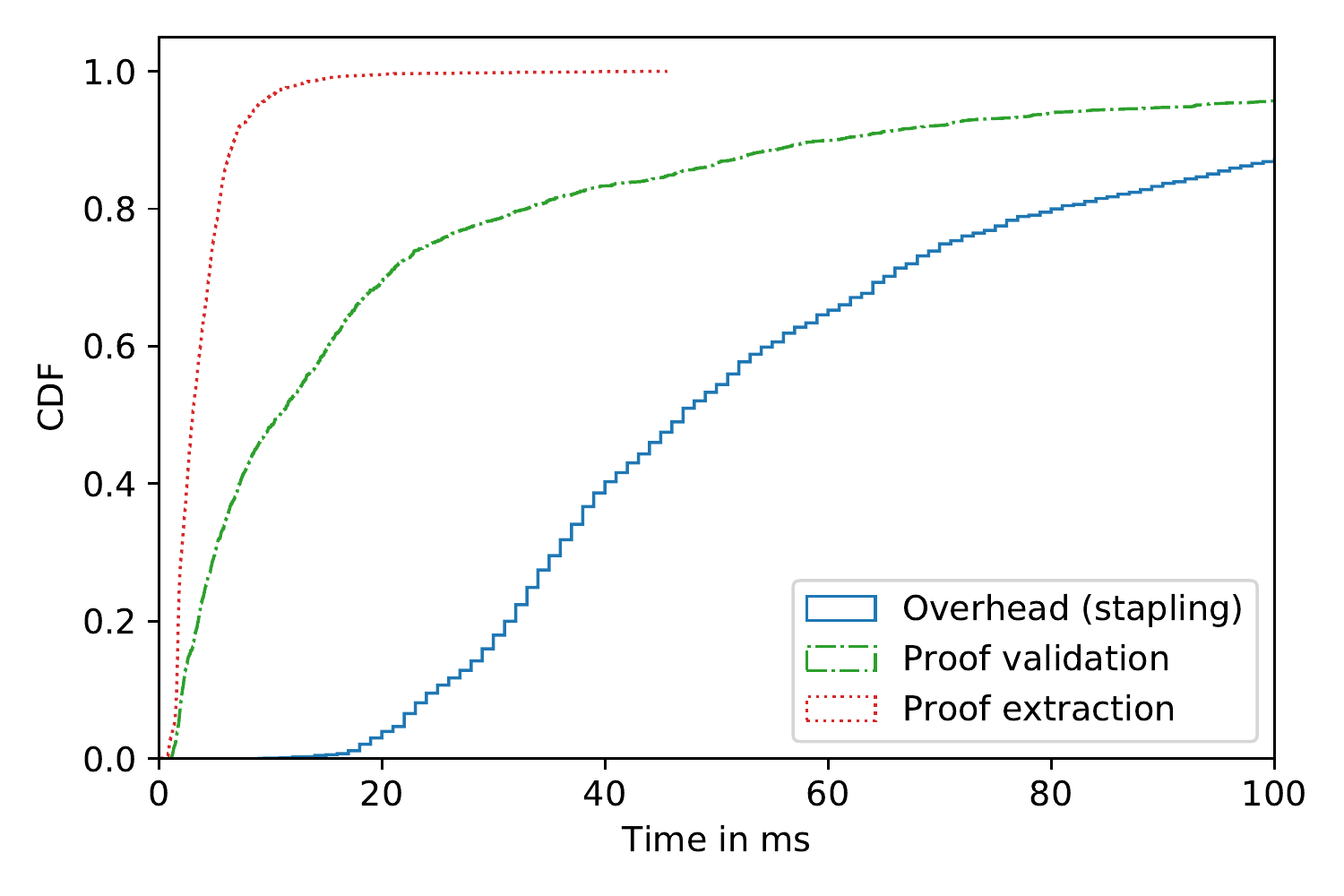}
  	\caption{Stapling approach: the proof is extracted from a TLS extension. Overhead is the
  	addition of extraction, validation, and extra latency incurred by our implementation as a
  	browser extension.}
  	\label{fig:e2e-stapling}
  \end{subfigure}
  \begin{subfigure}{\columnwidth}
  	\centering
  	\includegraphics[width=.8\linewidth]{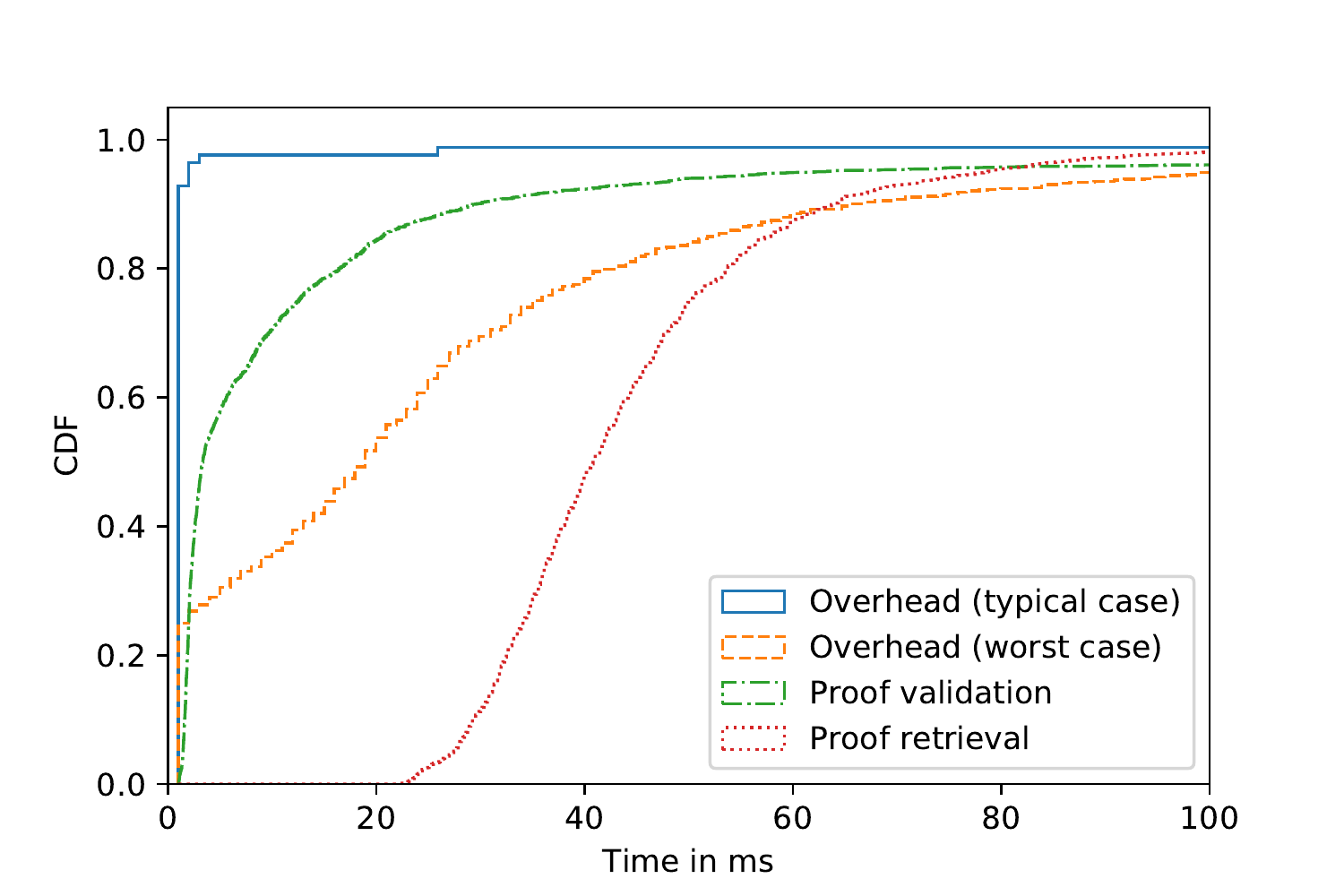}
  	\caption{DNS approach: the proof is requested via DNS from the map server. Typical case:
  	Complete DNS resolution happens in parallel. Worst case: IP address of the web server is cached
  	but not the proof.}
  	\label{fig:e2e}
  \end{subfigure}
  \caption{Latency incurred by \name operations. Overhead refers to the increase in TTFB, i.e.,
  from the moment the user requests a website to the moment it starts loading.}
  \label{fig:latency}
\end{figure}

\paragraph{Stapling.}
\Cref{fig:e2e-stapling} shows the TTFB overhead using the stapling approach. The processing time for
validating the proof has a median of $\sim$\SI{10}{ms}. However, the use of a JavaScript-based web
extension, instead of natively integrating the validation in the browser, leads to a TTFB increase
of around \SI{50}{ms}. A native browser implementation would reduce the overhead of both the
stapling and DNS methods.

\paragraph{DNS.}
To evaluate the DNS approach, we used the CoreDNS plugin to serve map server proofs as DNS
responses, as described in \cref{sec:proof-delivery}. We measured the performance with and without
locally caching the IP address of the web server. \Cref{fig:e2e} shows that fetching a newly
generated proof from the map server typically takes less than \SI{50}{ms}, while validating it
typically takes less than \SI{10}{ms}. In the typical case where neither IP addresses nor proofs are
cached, there is almost no impact on user experience ($<\SI{1}{ms}$ overhead in over 95\% of cases).
The overhead is negligible since the proof retrieval and validation, which are initiated in parallel
when the user requests a website, are often faster than the browser's regular workflow, i.e.,
DNS$\rightarrow$TCP/QUIC$\rightarrow$TLS$\rightarrow$HTTP\@.  For the second case, the IP address is
cached in the DNS resolver while the proof from the map server must be fetched.  Even in this worst
case, a median overhead of $\sim$\SI{20}{ms} means \name{} overhead is typically not noticeable.

\begin{figure}[h]
  \centering
  \includegraphics[width=.8\linewidth]{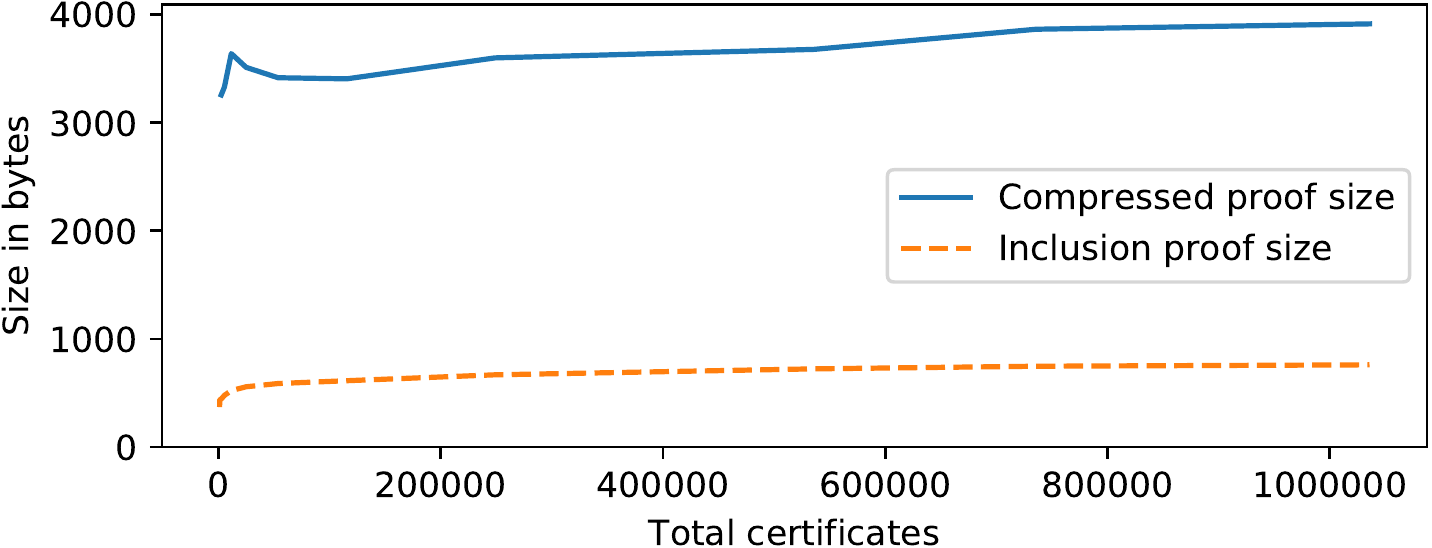}
  \vspace{-1ex}
  \caption{Average sizes of a complete proof (including the set of certificates, compressed) and
    an inclusion proof.}
  \label{fig:proofsize-over-time}
\end{figure}

\paragraph{Proof Size.}
\Cref{fig:proofsize-over-time} shows the average proof size (including the inclusion proof) after
DEFLATE compression and the size of the hash chains required to prove inclusion or non-inclusion in
the Merkle hash trees. We see that the inclusion proofs consisting of MHT hash chains contribute a
small, but incompressible part to the proof.  As expected, the proof size grows logarithmically with
the number of certificates/proofs. In the DNS proof retrieval method, we used the EDNS(0) extension
to increase the DNS payload size (DNS over TCP, TLS, or HTTPS should be used for larger proofs).

\begin{figure}[h]
  \centering
  \includegraphics[width=.8\linewidth]{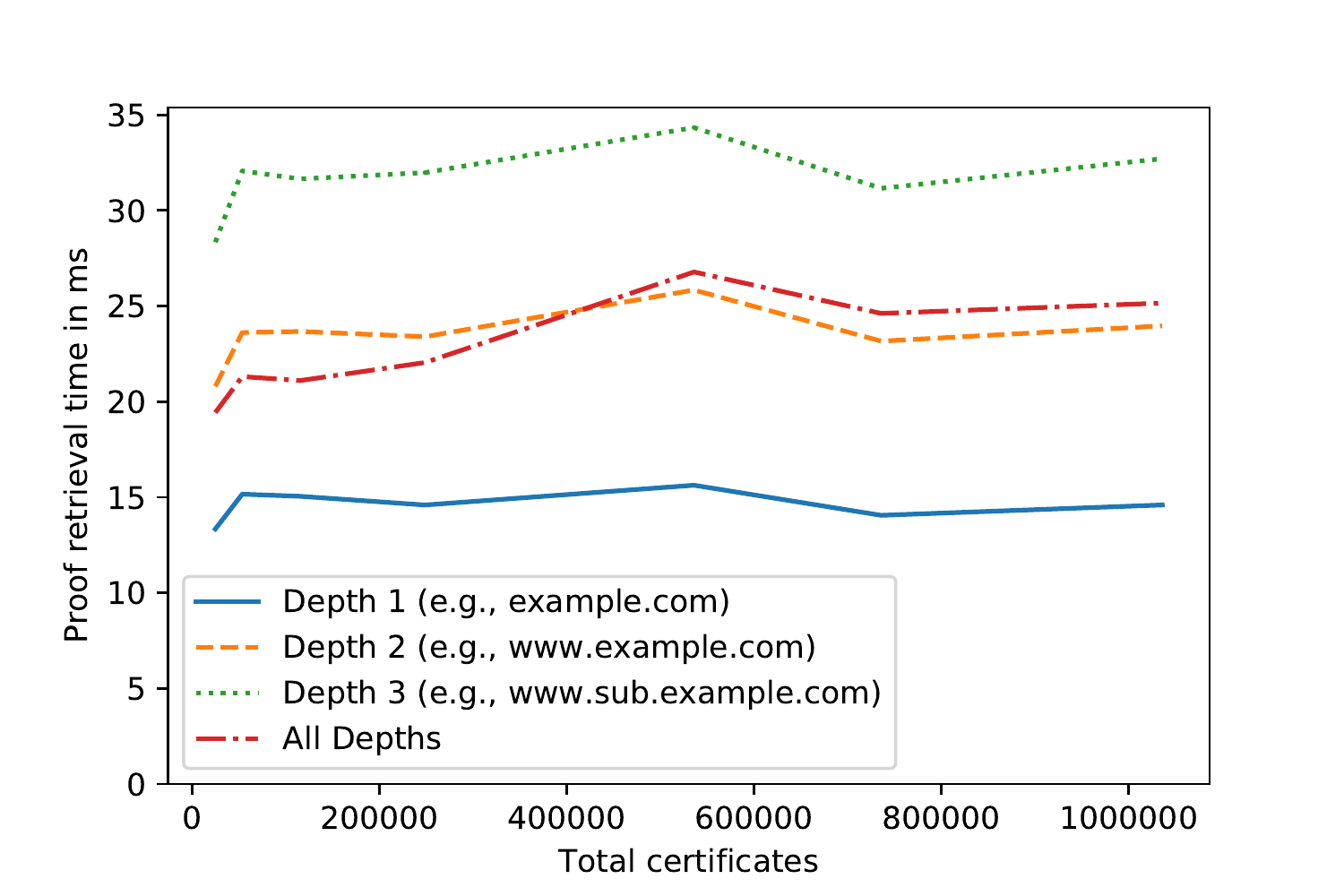}
  \caption{The average processing time of a map server for generating proofs given the total number
    of certificates added to the map server and the number of subdomains.}
  \label{fig:retrieval-time}
\end{figure}

\paragraph{Proof Generation.}
\Cref{fig:retrieval-time} shows that the average proof generation time of a map server increases
linearly with the number of subdomains due to an additional lookup in a sparse MHT per subdomain.
The generation time is almost constant given the number of certificates added to the map server. The
increase in generation time for all depths is due to an increase in the domain name depth of
certificates fetched from the CT log.  The virtual machine hosting the map server has 4 cores and
\SI{8}{\giga\byte} RAM\@. The proof-of-concept implementation creates a fully functional Trillian
sparse MHT for each domain with subdomains, even if there is only a single subdomain. In a leaner
implementation, MHTs for subdomains with few certificates could be generated on-the-fly. It is
important to note that within the maximum merge delay (MMD) of a map server, the map server can
precompute the proofs, which can already be cached in DNS resolvers.

\paragraph{Number of Subdomains.}
We fetched certificates from all Google-operated CT logs to analyze the overhead at map servers for
generating proofs. Fetching from all Google-operated CT log servers would (eventually) provide us
with all Chrome-accepted certificates since Chrome's CT policy mandates that each certificate be
logged by at least one Google-operated log server. The hierarchy of \name{} requires one additional
MHT inclusion proof per subdomain before the public suffix (e.g., .com or .co.uk). We fetched over
20 million certificates and stored the number of subdomains for the certificate's subject common
name (CN) or a subject alternative name (SAN) if the subject common name was empty. Our analysis
shows that most certificates (about $78\%$) have just one or two subdomain levels, which limits
database access to at most two queries for fetching an entry's proof (assuming cached public suffix
entries) and three queries for inserting an entry. Virtually all certificates have less than 6
subdomain levels.

\subsection{Deployment}

\name{} does not require server updates and only requires CAs to support an additional certificate
extension. This facilitates incremental deployment, which is crucial for the success of a proposed
infrastructure. \name{} could be initially deployed with a single map server and then grow to have
clients query a larger quorum of map servers. However, for scalability reasons and to distribute the
workload, it is possible---and would be sensible---to deploy separate map servers for different
subsets of the DNS namespace (for example, having a map server for each public suffix or TLD). Our
analysis of certificates contained in Google-operated CT logs indicates that the .com TLD appears in
54.77\% of all certificates, whereas the second most common TLD (.net) only appears in 7.29\% of
certificates. Therefore, the .com TLD should further be split to more evenly spread the workload
across map servers. Moreover, as we noted in \Cref{sec:server-selection}, different map servers can
support different sets of CAs.

In an initial phase of deployment, we also envision that map servers will serve as verifiable
monitors, without requiring clients to contact them directly and without enforcing domain policies.
Indeed, CT monitors have proved to be unreliable~\cite{li2019certificate}. Making monitors
verifiable with the data structures presented for map server would bring even more transparency to
the current ecosystem, before \name is fully operational. In later stages, domain owners would start
defining policies, before browsers enforce them.

\section{Discussion}

In this section, we discuss issues that relate to how \name{} would be used, deployed, and
configured in practice.

\subsection{Soft Fail vs. Hard Fail}

\name{} clients must obtain a global view of certificates for a domain from map servers. While this
is a conceptual change from how certificates are validated today, such a check is
necessary for enforcing strong security policies. Revocation has identical requirements: revocation
messages must be obtained---periodically or in real time---from a third party. Revocation messages
can be fetched by the client or stapled to the connection by the server. The same can be said of
data coming from our map servers. This is inevitable and may come at odds with availability,
especially in a world where web servers are rarely updated to support new security enhancements.

Revocation mechanisms (such as OCSP) are often implemented by browsers in a soft-fail
mode: if no revocation status can be obtained, the browser assumes that the certificate is not
revoked. Adam Langley compares soft-fail checks to ``a seat-belt that snaps when you
crash''~\cite{langley_crlsets}. With a soft-fail approach, previously received certificates should
be cached (until expired) so that the policies they contain cannot be downgraded by an attacker
capable of blocking connections to map servers. This approach is similar to HSTS~\cite{rfc6797},
OCSP Must-Staple~\cite{rfc7633}, Expect-CT~\cite{ietf-httpbis-expect-ct-08}, and HPKP~\cite{HPKP},
in that extra policies must be enforced after the first connection. In hard-fail mode (i.e., when
the connection to the web server is completely blocked by the browser because of missing
certificate-related data), relying parties must make sure that their trusted map servers are highly
available and redundant, to avoid breaking legitimate connections when a single map server goes
offline.

This discussion relates to the so-called ``criticality'' of X.509v3 extensions; a
non-critical certificate extension may be ignored if the browser does not recognize it, but must
be processed if it does. Our extensions should thus be defined as critical only when \name is
fully operational and a vast majority of browsers support it.

\subsection{Configuring Validation Policies}

Validation policies must support trust heterogeneity while providing sensible default values and an
intuitive interface for users to modify their policies. Operating system and browser vendors, the
CA/Browser Forum, and other organizations such as EFF, ICANN, and Mozilla could provide validation
policy packages that are either installed by default or can be fetched, possibly modified, and
installed by users.

Consider three users with different awareness for security: Alice who does not have strong security
awareness, Bob who is aware of the importance of security but is not particularly tech-savvy, and
Charlie who is an IT expert who is able and willing to customize his setup. Let $B$ be a browser
that supports \name{}. Alice installs $B$ and does not modify default trust levels. Even such a
general validation policy would improve the security compared to today's web PKI, as it prevents
downgrade attacks to HTTP. Indeed, when Alice visits a plain HTTP website, her browser can verify
with \name that she's not under attack by checking the certificates and policies relevant to the
domain name in question (see \cref{sec:http_downgrade} for more details). Bob installs $B$ and a
validation policy package that fits his needs. Such a policy package could define as highly trusted
CAs that support multi-perspective validation, which prevents BGP hijacking attacks, or national
(e.g., American) CAs for websites with a national TLD (e.g., .us and .gov) to prevent attacks by a
foreign CA, see \cref{sec:valid-polic-trust} for more examples. Charlie installs $B$ and a
validation policy package, then adjusts the validation policy, e.g., by adding or removing CAs from
the set of highly trusted CAs to provide fine-grained control over the trust decisions. F-PKI thus
provides flexibility and tangible benefits to all users.

\subsection{HTTP-Downgrading Prevention}
\label{sec:http_downgrade}

\name{} has the extra benefit of allowing clients to efficiently verify, when they connect to a
plain HTTP website, that no valid certificates currently exist for the corresponding domain,
eliminating any possibility of a downgrade attack. If a user visits such a website using the \name{}
browser extension, the extension detects any downgrade attack and redirects the user or blocks the
connection. With such a system in place, it becomes possible to both tolerate regular HTTP websites,
and block websites that exhibit the characteristics of a downgrade attack. This shows that it is
possible to support a legacy protocol without deteriorating the security of a new protocol. HTTP
Strict Transport Security (HSTS) aims at solving the same problem, but it must be configured by the
website administrator and has limitations as discussed in \cref{sec:lessons-learned}.

\subsection{On the Multi-Certificate Approach}

Following the observation that a PKI should \emph{prevent} a single CA from issuing illegitimate
certificates for any websites, previous work has made extended use of multi-signed
certificates~\cite{Wang2015CMS,Basin2014,Szala2014Poli,syta2015certificate}. Regrettably, requiring
signatures from multiple CAs has drawbacks as well. First, it increases the complexity of the
system, particularly if CAs must communicate and coordinate with each other, which is not the case
in traditional PKIs. Second, an inherent tradeoff lies in the number of signatures: requiring more
entities to certify the same public key, although it increases resilience, may be more expensive and
deteriorates performance and usability. Finally, consider a government launching a surveillance
campaign through man-in-the-middle attacks: requiring that certificates be multi-signed does not
prevent any attacks if the country in question controls a sufficient number of CAs.

\name{} allows domain owners to use multiple certificates for the same domain name, but gives more
flexibility to domain owners without compromising on security.  Although our PKI offers a set of
strong security guarantees even for single certificates, multi-certificates could be considered an
additional assurance of the binding between public key and name. Browsers could even display a
special security indicator when a public key is certified by independent CAs, without making
multi-certificates mandatory for all domains.

\section{Related Work}
\label{sec:related}

\name{} presents similarities with the Perspectives system proposed by Wendlandt et
al.~\cite{wendlandt2008perspectives}, in that both aim at giving clients a broader view of existing
public keys (to detect illegitimate keys) by relying on a third party. Perspectives allows SSH or
HTTPS clients to detect man-in-the-middle attacks by  contacting ``semi-trusted'' hosts called
network notaries. We improve upon that approach by using verifiable data structures to reduce the
trust put in those third parties and enable an efficient detection of logging inconsistencies.

Among the most comprehensive proposals for redesigning the web PKI from the ground up are
AKI~\cite{KimHuaPerJacGli13} and its successor ARPKI~\cite{Basin2014}. Both systems are based on
multi-signed certificates, which could be used to enhance the security of \name even further.
ARPKI provides resilience to the compromise of up to $n-1$ entities,
where $n \ge 3$ is a system parameter. PoliCert~\cite{Szala2014Poli} builds upon ARPKI to let each
domain owner define its own security parameters and policies. PoliCert introduces policies similar
to ours, but they are defined in a separate policy file (called SCP) rather than in the certificate
as we do here. This file again relies on a multi-signing approach to guarantee the authenticity of
the policies it defines. SCPs allow domain owners to specify which CAs they trust, but the approach
suffers from a circular dependency as SCP themselves must be signed by CAs, without a guarantee that
all relevant policies will always be delivered to the client. DTKI~\cite{yu2016dtki} allows domain
owners to define a master certificate which signs its TLS certificates. This master certificate is
added to a public append-only log using MHTs to provide a comprehensive set of valid and invalid
certificates.  Neither PoliCert nor DTKI support heterogenous trust on the client side: all CAs are
equally trusted.

Syta et al.~\cite{syta2015certificate} proposed to address the weakest-link security problem of
today's infrastructure by deploying a single ``Cothorithy'' that federates all existing CAs. This
cothorithy would use CoSi, a collective signing protocol that scales to a large number of signers,
to allow each CA to detect fake certificates before they are signed. However, the authors do not
provide a detailed description of how these certificates would be detected and blocked in
practice~\cite{syta2015certificate}. One of the challenges of multi-signed certificates is in
defining the right threshold number of required signatures. Moreover, it is unclear whether
multi-signed certificates would prevent state-level adversaries that control a large number of CAs
from issuing illegitimate certificates.

The trust management system by Braun~et~al.~\cite{ca-trust-management} fixes issues in the web PKI
and also defines trust levels. However, their approach is different from \name{}, as they reduce the
attack surface (trusted CAs) based on local knowledge and reputation, without proofs of
presence/absence of policies and other certificates.

Certificate revocation should be an integral part of any PKI design, but it is a vast and active
research area in and of itself~\cite{SoKDeleg2020}. Several revocation schemes have been proposed
in academic literature in the last few years, but browser vendors have also developed and deployed
their own schemes~\cite{onecrl,langley_crlsets}.
Although \name has a built-in revocation mechanism, it could be combined with (or extended by)
stand-alone revocation schemes to improve efficiency or achieve additional security properties.
CRLite~\cite{larisch2017crlite}, for example, uses Bloom filters to efficiently distribute
revocation messages to clients. \name{} has similar objectives as
CRLite~\cite[Section~4]{larisch2017crlite} but membership lookup is not sufficient for \name{}
because clients need to obtain a comprehensive set of domain policies. PKISN~\cite{PKISN2016}
is also a log-based revocation system but focuses on a specific problem: that of
revoking CA certificates without causing collateral damage.

Dahlberg et al.~\cite{Dahlberg2018} proposed light-weight monitoring (LWM), which allows domain
owners to perform self-monitoring to detect bogus certificates issued for their domain via
untrusted notifiers. LWM uses a MHT data structure based on lexicographically ordered domains
similar to our sorted list MHT discussed in \cref{sec:denial-of-service-vectors-and-mitigations}.
This data structure enables efficient generation of inclusion and non-inclusion proofs for
certificates. The main difference to \name{} is that LWM requires constant self-monitoring by the
domain owner to detect bogus certificates, requires modifications at the CT log servers, and
provides no proactive security guarantees for users.

\section{Conclusion}

The web PKI should not only be equipped with means of detecting illegitimate certificates but it
should also prevent these certificates from being used in attacks altogether.
\name meets this objective while taking into account the challenges that come with deploying such
a scheme.
\name{} does not require active CA participation, is backward compatible,
supports multiple deployment models (including one that does not require updates on the server
side), and prevents impersonation attacks. Yet, it minimizes unintended breakage
by letting both domain owners and clients define their own policies. Our prototype implementation,
requiring only a browser extension and a map server, demonstrates that such a system can be
deployed with minimal changes to the current infrastructure.

Trust is a complex and heterogeneous notion that is only partially captured by traditional PKIs
where CAs are omnipotent. We propose a more flexible model in which authenticity is derived not only
from a certificate chain but also from a global view of signed statements. With such a view, relying
parties can make informed decisions and only consider a certificate authentic when no conflicting
statements have been issued by an authority considered highly trusted for the domain name in
question.

The map servers we introduce in this paper are used to overcome the deficiencies of CT's log servers
without replacing them: they make it possible to efficiently and verifiably obtain all the data
relevant to any domain, for monitoring purposes and for achieving the additional security properties
we presented. The data structure we propose for map servers enables clients to efficiently detect
policy violations and inconsistencies.

\section*{Acknowledgments}
We thank the anonymous reviewers, our shepherd, Samuel Jero, and Paul Syverson for their highly
valuable feedback.

This project has received funding from the European Union's Horizon 2020 research and innovation
programme under grant agreements No 825310 and 825322.
We gratefully acknowledge support from ETH Zurich, from the Zurich Information Security and Privacy
Center (ZISC), and from the WSS Centre for Cyber Trust at ETH Zurich.

\bibliographystyle{IEEEtranS}
\bibliography{references.bib}

\appendix

\subsection{Abstract Model of \name{}}
\label{sec:formal_model}

The abstract PKI model presented herein is based on the model introduced by
Maurer~\cite{maurer1996modelling} and later extended by Marchesini et
al.~\cite{marchesini2005modeling}. We extend it further, primarily to allow relying parties to
request policies that cover a set of highly trusted CAs (which is a concept distinct from Maurer's
recommendation levels or confidence values).

The \emph{view} of a relying party in the public-key infrastructure is modeled as a set of
statements. A statement is valid if it is either part of an initial view (axiomatic) or derived from
a view using one of the inference rules we present below. Statements can take one of the following
forms:

\begin{itemize}
	\item \textbf{Authenticity of Binding:} $\textit{Aut}(X, N, \mathcal{R}, \mathcal{I})$ denotes
	the belief that public key $X$ is bound to name~$N$ and trustworthy for issuing certificates
	over realm~$\mathcal{R}$ during time interval $\mathcal{I}$. The issuance realm $\mathcal{R}$
	is the set of names for which the public key $X$ is considered authoritative.
	\item \textbf{Certificate:} $\textit{Cert}(X_1, X_2, N, \mathcal{R}, \mathcal{I})$ indicates
	that the owner of public key $X_1$ has issued a certificate to the owner of $X_2$, which binds
	public key $X_2$ to name~$N$ and attests that the owner of $X_2$ is trustworthy for issuing
	certificates over realm~$\mathcal{R}$. This certificate is valid during interval~$\mathcal{I}$.
	\item \textbf{Log Trust:} $\textit{LogTrust}(L, \mathcal{S})$ denotes trust in log $L$ for
	recording certificates issued by authorities in set $\mathcal{S}$.
	\item \textbf{Proof of Compliance:} $\textit{Proof}(L, X, N, \mathcal{I})$ denotes a proof
	coming from log $L$ that public key $X$ complies with policies defined for name $N$. This
	proof is valid during interval $\mathcal{I}$.
	\item \textbf{Compliant:} $\textit{Compliant}(X, N, \mathcal{C}, \mathcal{I})$ denotes the
	belief that public key $X$ is compliant with all policies defined for name $N$ by entities in
	set $\mathcal{C}$ during interval $\mathcal{I}$.
\end{itemize}

A proof of compliance is to a \textit{``Compliant''} statement what a certificate is to an
authenticity statement: a cryptographic object used by the relying party to derive an actual belief,
for a limited amount of time, under the condition that the entity that produced the cryptographic
object is trusted for doing so. A compliance statement can thus be derived from a proof of
compliance and a log-trust statement as follows:

\begin{equation}
  \begin{gathered}
    \textit{LogTrust}(L, \mathcal{S}), \\
    \textit{Proof}(L, X, N, \mathcal{I}) \vdash \\[1ex]
	\textit{Compliant}(X, N, \mathcal{S}, \mathcal{I})
  \end{gathered}
\end{equation}

Two compliance statements can be combined, in which case the resulting statement covers the union of
both CA sets, but is only valid for the intersection of the validity intervals:

\begin{equation}
  \begin{gathered}
    \textit{Compliant}(X, N, \mathcal{S}_1, \mathcal{I}_1), \\
    \textit{Compliant}(X', N, \mathcal{S}_2, \mathcal{I}_2) \vdash \\[1ex]
	\textit{Compliant}(X, N, \mathcal{S}_1 \cup \mathcal{S}_2, \mathcal{I}_1 \cap \mathcal{I}_2)
  \end{gathered}
\end{equation}

If no valid certificate exists for name $N$, then the log server may produce a proof where public
key $X$ is null (i.e., $X=\emptyset$). For this reason, in the above equation, $X'$ may be equal to
either $X$ or $\emptyset$.

The new trust level is introduced with the following function:

\begin{center}
	$f(N)$: set of authorities highly trusted for name $N$
\end{center}

Authenticity is then derived from the following:
(a) an authenticity statement, which specifies the realm $R_1$ over which the issuer has authority;
(b) a certificate statement, declaring the subject's name $N_2 \in R_1$; and
(c) a compliance statement for that name such that
$f(N_2) \subseteq \mathcal{S}$:
\begin{equation}
  \begin{gathered}
    \textit{Aut}(X_1, N_1, \mathcal{R}_1, \mathcal{I}_1), \\
    \textit{Cert}(X_1, X_2, N_2, \mathcal{R}_2, \mathcal{I}_2), \\
    \textit{Compliant}(X_2, N_2, \mathcal{S}, \mathcal{I}_3) \vdash \\[1ex]
	\textit{Aut}(X_2, N_2, \mathcal{R}_1 \cap \mathcal{R}_2, \mathcal{I}_1 \cap \mathcal{I}_2 \cap
        \mathcal{I}_3), \\
  \end{gathered}
\end{equation}

In contrast to previous models, we do not use a dedicated $\textit{Trust}$ statement for CAs, to
model the concept of ``trustworthiness for issuing certificates''~\cite{maurer1996modelling}.
Instead, we include the issuance \emph{realm} into authenticity statements. This realm would
typically be the set of all possible names for CAs (with no name constraints) and the empty set for
non-CA entities. This allows us to bind that realm with the rest of the authenticity statement, and
more accurately represent the HTTPS public-key infrastructure as parameters that relate to that
issuance realm (the CA bit, name constraints) are defined in the certificate together with other
parameters such as the validity period. However, we employ a separate $\textit{LogTrust}$ statement
to model entries in the list of trusted logs that relying parties must establish to support any
log-based scheme.

\subsection{Construction of the MHT}
\label{sec:map_server_construction}

Certificates and revocations are added to the map server as follows. First the
\textsc{ConstructDictionary} method in \Cref{alg:mapserverconstruction} constructs a hierarchical
dictionary structure that maps domain names to their corresponding certificates and revocations. The
root dictionary, denoted by $R$, contains entries for all e2LDs, and each of these entries may
contain a dictionary for its subdomains, whose entries can in turn have further subdomain
dictionaries, forming a hierarchical data structure with one dictionary level per nested subdomain.
Each certificate and revocation is added to every domain listed in either the common name (CN)
attribute of the subject field or as dNSName in the subject alternative name (SAN) extension of the
(revoked) certificate. After creating $R$, the MHT structures are created or updated in a bottom-up
fashion, starting with the subdomain sparse MHTs and finishing with the e2LD sparse MHT root.

\begin{algorithm}
  \small
  \DontPrintSemicolon
  \SetKwIF{If}{ElseIf}{Else}{if}{}{else if}{else}{end if}
  \SetKwProg{function}{function}{}{}
  \SetKwRepeat{Struct}{struct \{}{\}}%
  $L$: list of certificates and revocation messages \\
  \Struct{Entry}{\KwSty{list} certificates\\
    \KwSty{list} revocations\\
    \KwSty{dictionary} subdomains}
  $R$: dictionary (key: \textit{domain}; value: \textit{Entry})\\
  \function{\textsc{ConstructDictionary}($L$)}{
    \For{$e \in L$}{
      \If{$\Call{IsRevocation}{e}$}{
        $c \gets e.\text{revoked\_certificate}$
      }\Else{
        $c \gets e$
      }
      \For{$d \in c.\textit{Subject} \cup c.\textit{SAN}$}{
        $d \gets \Call{RemoveWildcard}{d}$\\
        $A \gets R[\Call{E2LD}{d}]$ \\
        \For{$p \in \Call{ParentDomains}{d}$}{ \Comment{$p$ starts below the e2LD}\\
          $A \gets A.\text{subdomains}[p]$
        }
        \If{$\Call{IsRevocation}{e}$}{
          $\Call{Append}{A.\text{revocations}, e}$
        }\Else{
          $\Call{Append}{A.\text{certificates}, e}$
        }
      }
    }
    \Return $R$
  }
  \caption{Construction of the Map Server}
  \label{alg:mapserverconstruction}
\end{algorithm}

\subsection{Automatic Selection of Map Servers}
\label{sec:map_server_selection_algo}

Let $\mathcal{M}$ be the set of map servers, $\mathcal{C}$ the set of the client's highly trusted
CAs, and $Q$ the quorum of map servers that minimally need to support each CA in $\mathcal{C}$. For
a map server $M \in \mathcal{M}$, let $\textsc{Cost}(M)$ be the cost associated with map server $M$
and let $\textsc{Sup}(M)$ be the set of CAs supported by $M$. The task is to find $x_i$ for $1 \leq
i \leq |\mathcal{M}|$ which minimizes the total cost, where $x_M=1$ if map server $M$ is included in
the minimal set of map servers and $x_M=0$ otherwise.

\begin{equation*}
  \begin{array}{lllr}
    \text{minimize} & \displaystyle\sum_{M\in\mathcal{M}}\left\{
    \begin{array}{lr}
      \textsc{Cost}(M), & x_M=1\\
      0, & \text{otherwise}
    \end{array}\right. & &\\
    \text{subject to} & \displaystyle\sum_{M: c\in \textsc{Sup}(M)}x_M \geq Q, & c \in \mathcal{C} &\\
    & x_M \in \{0, 1\}, & M \in \mathcal{M} &\\
\end{array}
\end{equation*}

This is a set-multicover problem, which is NP-hard as we cannot assume that $\mathcal{M}$ has a
specific internal structure (e.g., Vapnik--Chervonenkis dimensions~\cite{chekuri2012set}). A greedy
algorithm can solve the set–multicover problem with time complexity
$O(|\mathcal{M}|\cdot{}{|\mathcal{C}|}^2\cdot{}Q)$. \Cref{alg:greedymulticover} describes this in
pseudo-code.

\begin{algorithm}
  \small
  \DontPrintSemicolon
  \SetKwIF{If}{ElseIf}{Else}{if}{}{else if}{else}{end if}
  \SetKwProg{function}{function}{}{}
  \SetKwRepeat{Struct}{struct \{}{\}}%
  $\mathcal{M}$: set of all map servers\\
  $M_{in}$: set of trusted map servers $M_{in} \subset \mathcal{M}$\\
  $\mathcal{C}$: set of all CAs\\
  $C_{in}$: set of highly trusted CAs $C_{in} \subset \mathcal{C}$\\
  $Q$: min. number of map servers for each CA in $C_{in}$\\
  $S$: multiset of trusted CAs covered by the map servers\\
  $\Call{Sup}{X}$: set of CAs supported by $X$, for $X \in \mathcal{M}$\\
  $\textsc{Alive}(X,S,Q) := |\{\forall s \in S: c = s\}| < Q$, returns true if CA $X \in \mathcal{C}$ is not yet sufficiently covered in $S$\\
  $\Call{N}{X,S,Q} := \displaystyle\sum_{c \in \Call{Sup}{X}}\left\{
    \begin{array}{lr}
      1, & \Call{alive}{c,S,Q}\\
      0, & \text{otherwise}
    \end{array}\right.$, returns the number of additional CAs covered by selecting map server $X \in \mathcal{M}$\\
  \function{\textsc{SelectMapServers}($M_{in}, C_{in}, Q$)}{
    $M_{out} \gets \Call{EmptySet}{}$\\
    $S \gets \Call{EmptyMultiset}{}$\\
    \While{$\exists c \in C_{in}: \Call{alive}{c,S,Q}$}{ \label{alg:line:loop-condition}
       \If{$\forall l \in M_{in}: \Call{N}{l,S,Q}=0$}{
         \Return \Call{EmptySet}{} \Comment{There is no multicover} \label{alg:line:no-solution}
       }
       $m_{opt} \gets \underset{m \in M_{in}}{\mathrm{argmin}} \frac{\Call{Cost}{m}}{\Call{N}{m,S,Q}}$ \Comment{Find best candidate} \label{alg:line:select-best-candidate}\\
       $M_{out} \gets M_{out} \cup m_{opt}$\\ \label{alg:line:add-selected-candidate}
       $M_{in} \gets M_{in} \setminus m_{opt}$\\ \label{alg:line:remove-selected-candidate}
       $C \gets C + \Call{Sup}{m_{opt}}$ \Comment{Add newly covered CAs}
    }
    \Return $M_{out}$
  }
  \caption{Greedy Algorithm for Set Multicover}
  \label{alg:greedymulticover}
\end{algorithm}

The algorithm sequentially selects the most cost-efficient map server
(\cref{alg:line:select-best-candidate}) until the quorum of map servers is achieved for all highly
trusted CAs (\cref{alg:line:loop-condition}). After adding a map server to the minimal set of map
servers (\cref{alg:line:add-selected-candidate}) it is removed from the pool of available map
servers to prevent picking the same map server more than once
(\cref{alg:line:remove-selected-candidate}). If none of the remaining map servers cover any of the
highly trusted CAs and the quorum is not yet achieved, there is no solution
(\cref{alg:line:no-solution}).

Rajagopalan et al.~\cite{rajagopalan1998primal} show that this algorithm returns a set multicover
with a cost of $C_\text{greedy}$ which compares to the optimal solution $C_\text{opt}$ as follows:
\begin{align}
  \label{eq:greedy-solution}
  C_\text{greedy} &= \sum_{e \in \mathcal{M}}\text{price}(e)\\
  C_\text{greedy} &\leq H_{|U|\cdot{}Q}\cdot{}C_\text{opt}, H_n = 1 + \frac{1}{2} + \frac{1}{3} + ... + \frac{1}{n}\\
  C_\text{greedy} &\leq (1+ln(|U|\cdot{}Q))\cdot{}C_\text{opt}
\end{align}

The algorithm allows each map server in $\mathcal{M}$ to be associated with a cost. Using a fixed
cost for all map servers minimizes the number of map servers.  Users could define Cost functions
related to trustworthiness of a map server or required network resources.  A map server might
support few CAs or restricts itself to a small subset of domains and thus produce smaller proofs
than large map servers.  Selecting several small map servers could result in a smaller overall proof
size.

\subsection{Denial-of-Service Vectors and Mitigations}\label{sec:denial-of-service-vectors-and-mitigations}

An attacker can attempt to launch a resource-exhaustion attack by injecting certificates for e2LDs
such that the hash of the injected domain and target domain share the same prefix. If an
attacker finds a domain such that the hash of this domain and the hash of the target domain have a
common prefix of length $l$ and there is no other domain in the sparse MHT whose hash has a common
prefix of length $l$ with the target domain, then the attacker can increase the proof size of the
target domain by 32 Bytes.

The number of prefix collisions an attacker can generate is bounded by the probability of finding a
collision $P(l_p=l)=1-{(1-2^{-l})}^m$ for each prefix length $l_p$ independently ($m$ is the number
of hash values calculated by the attacker).
We estimated that, if the attacker can perform $10^9$ hash calculations per second,
even after an attack period of one year, the proof size can only increase by 1100 bytes on average.

The first mitigation is to add a unique, unpredictable, and regularly updated tree identifier to
the map server and include this identifier in the node, leaf, and domain hash computation of the
sparse MHT, similar to the tree-wide nonce $k_n$ used in CONIKS~\cite[Chapter~3.1]{melara2015coniks}.

\begin{figure}[t]
  \centering
  \resizebox{0.5\textwidth}{!}{%
    \input{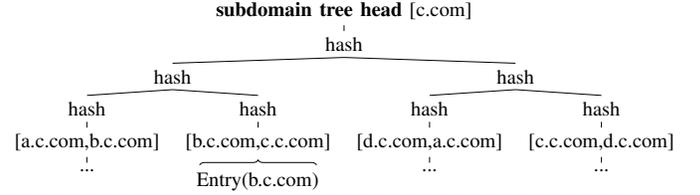}}
  \caption{Sorted-list MHT, as an alternative to a sparse Merkle tree, containing subdomains of c.com.}
  \label{fig:hierarchical_subdomain_lists}
\end{figure}

The second mitigation is to use a different data structure for verifiable logging that produces
constant-sized proofs. A sorted-list MHT~\cite{RevTrans} is an MHT where each leaf points to the
domain lexicographically adjacent to its own. A leaf thus consists of its domain $d_1$ and the
corresponding entry, and the next domain $d_2$, such that there is no domain $d$ with $d_1<d<d_2$.
\Cref{fig:hierarchical_subdomain_lists} shows a sorted-list MHT for the subdomains of
\textit{c.com}. Leaves need not be stored in a specific order as long as each leaf points to the
adjacent domain and the domain pairs form a cycle which contains all domains. The hash computation
of leaves and intermediate nodes is the same as in a sparse MHT, explained in
\cref{sec:verifiable_logging}, except that a leaf's value additionally contains the DER
representation of $d_1$ and $d_2$ between the fixed leaf prefix and the DER representation of the
entry. Existence of domain $d$ is proven by providing the hash inclusion proof for the leaf where
$d=d_1$. Non-existence of domain $d$ is proven by the inclusion proof for the leaf with the domain
pair $d_1, d_2$ such that $d_1<d<d_2$.  Proofs in a sorted-list MHT have the same size as regular
MHT inclusion proofs ($\ceil*{\text{log}_2L}$) in addition to the adjacent domain which is included
in the proof.  In a sorted-list MHT, updating an entry requires changing either
$\ceil*{\text{log}_2L}$ or $\ceil*{\text{log}_2L}-1$ node values. Adding and removing entries can,
depending on the tree structure and the inserted domain, in the worst case require changing up to
$3\cdot{}\ceil*{\text{log}_2L}$ node values. While both sparse MHTs and sorted-list MHTs provide a
verifiable map based on a verifiable log, sparse MHTs produce larger proofs on average but can be
updated more efficiently.

Both mitigations incur a
performance penalty either through less efficient update operations on sorted-list MHTs or through
the necessity of rebuilding the complete tree when changing the tree identifier.

\end{document}